\documentclass[12pt]{iopart}
\usepackage{graphicx}
\eqnobysec

\newcommand{\beq}{\begin{equation}}
\newcommand{\beqa}{\begin{eqnarray}}
\newcommand{\eeq}{\end{equation}}
\newcommand{\eeqa}{\end{eqnarray}}
\newcommand{\bfa}{{\bf a}}
\newcommand{\bfb}{{\bf b}}
\newcommand{\bfd}{{\bf d}}
\newcommand{\bfe}{{\bf e}}
\newcommand{\bfBB}{{\hskip -.3mm}{\bf B{\hskip -.6mm}B}{\hskip -.3mm}}
\newcommand{\bfSS}{{\hskip -.1mm}{\bf S{\hskip -.1mm}S}{\hskip -.1mm}}
\newcommand{\bg}{{\rm bg}}
\newcommand{\bigmean}[1]{\left\langle#1\right\rangle}
\renewcommand{\d}{{\rm d}}
\newcommand{\dr}{\delta\rho}
\newcommand{\ds}{\displaystyle}
\renewcommand{\e}{{\rm e}}
\newcommand{\frad}[2]{\displaystyle{\displaystyle#1\over\displaystyle#2}}
\newcommand{\gl}{{\rm gl}}
\newcommand{\hdr}{\widehat{\delta\rho}}
\renewcommand{\i}{{\rm i}}
\newcommand{\loc}{{\rm loc}}
\renewcommand{\max}{{\rm max}}
\newcommand{\mean}[1]{\langle#1\rangle}
\newcommand{\prob}{\mathop{\rm Prob}\nolimits}
\newcommand{\BB}{{\rm BB}}
\newcommand{\BS}{{\rm BS}}
\newcommand{\J}{{\bf J}}
\newcommand{\N}{{\cal N}}
\newcommand{\SB}{{\rm SB}}
\renewcommand{\SS}{{\rm SS}}

\begin{document}

\title{Structure of the stationary state of the asymmetric target process}

\author{J M Luck and C Godr\`eche}

\address{Service de Physique Th\'eorique\footnote{URA 2306 of CNRS},
CEA Saclay, 91191 Gif-sur-Yvette cedex, France}

\begin{abstract}
We introduce a novel migration process, the target process.
This process is dual to the zero-range process (ZRP) in the sense that,
while for the ZRP the rate of transfer of a particle
only depends on the occupation of the departure site,
it only depends on the occupation of the arrival site for the target process.
More precisely, duality associates to a given ZRP a unique target process,
and vice-versa.
If the dynamics is symmetric, i.e., in the absence of a bias,
both processes have the same stationary-state product measure.
In this work we focus our interest on the situation where
the latter measure exhibits a continuous condensation transition
at some finite critical density $\rho_c$, irrespective of the dimensionality.
The novelty comes from the case of asymmetric dynamics,
where the target process has a nontrivial fluctuating stationary state,
whose characteristics depend on the dimensionality.
In one dimension, the system remains homogeneous at any finite density.
An alternating scenario however prevails in the high-density regime:
typical configurations consist of long
alternating sequences of highly occupied and less occupied sites.
The local density of the latter is equal to~$\rho_c$
and their occupation distribution is critical.
In dimension two and above, the asymmetric target process exhibits a phase
transition at a threshold density $\rho_0$ much larger than~$\rho_c$.
The system is homogeneous at any density below $\rho_0$,
whereas for higher densities it exhibits an extended condensate
elongated along the direction of the mean current,
on top of a critical background with density $\rho_c$.
\end{abstract}


\eads{\mailto{jean-marc.luck@cea.fr},
\mailto{claude.godreche@cea.fr}}

\maketitle

\section{Introduction}

In recent years many studies have been devoted to
nonequilibrium statistical-mechanical models yielding condensation,
such as zero-range processes
(ZRP)~\cite{loan,evans1,wis1,cg,gross,evans2,glcond},
dynamical urn models~\cite{zeta1,zeta2,zeta3,barc,lux},
and mass transport models~\cite{maj}.
In all these models the condensate manifests itself by
the macroscopic occupation of a single site
by a finite fraction of the whole available mass.

The ZRP is the simplest of these models.
It is a driven diffusive system with multiple occupations,
such that particles hop from site to site on a lattice,
with a rate which only depends on the occupation of the departure site.
The distribution of the particles among the sites
in the stationary state is given by a product measure,
which is explicitly known in terms of the rate defining the model,
irrespective of the geometry of the system
and of the asymmetry of the dynamics~\cite{evans2,spitz,andj}.
This property of the stationary-state measure of the ZRP
favors the condensation phenomenon.
The product structure indeed prevents the condensate
from being shared by more than one site~\cite{gross,glcond,maj}.

Dynamical urn models, also called migration processes in the probabilistic
li\-te\-ra\-ture~\cite{kelly}, can be viewed as generalizations of the ZRP,
where the rate at which a particle is transferred
from a departure site to an arrival site
now depends on the occupations of both sites.
Migration processes do not have a stationary-state product measure in general.

In the present work we introduce a special migration process,
the {\it target process}.
This process is novel to our knowledge.
It has no stationary-state product measure, except in the case
of a symmetric dynamics, i.e., in the absence of a bias,
leading to an equilibrium stationary state.
The class of target processes is dual to the class of ZRP,
in the sense that the roles of the departure and arrival sites are interchanged:
the rate basically depends on the occupation of the departure (source) site
for a ZRP,
and on the occupation of the arrival (target) site for a target process.
More precisely, to a given ZRP is associated by duality a unique target process,
and vice-versa.
In the case of a symmetric dynamics,
these two dual processes have the same stationary-state product measure.

Our aim is to study the structure of the nonequilibrium stationary state
of the asymmetric target process,
and especially the fate of the condensation phenomenon.
In Section~\ref{stp} we give a detailed definition
of the class of target processes.
We then focus our attention onto the particular target process
which is dual to the `canonical ZRP'
studied e.g.~in~\cite{evans1,wis1,cg,gross}.
In the case of symmetric dynamics, the target process thus constructed
has the same stationary-state product measure as the ZRP,
and therefore the same continuous condensation transition.
In the asymmetric case, however, the stationary-state measure of the target
process is not a product measure,
and exhibits non-trivial correlations in general.
This basic difference between the ZRP and the target process
manifests itself more drastically at high density and in low dimensionality.
Sections~\ref{1D1} and~\ref{1D2} are devoted to a thorough theoretical
and numerical study of the one-dimensional situation.
The system remains homogeneous at any finite density,
and presents an alternating structure which is more and more pronounced
as the density is increased.
The two-dimensional asymmetric target process
on the square lattice is the subject of Section~\ref{2D}.
It exhibits an unconventional type of condensation phenomenon,
with a transition at a threshold density $\rho_0$ much larger than~$\rho_c$,
and an extended condensate elongated along the direction of the bias.
An analogous scenario presumably generically holds on other lattices
and in higher dimension as well.
Section~\ref{quatre} contains a Discussion.

\section{The target process}
\label{stp}

\subsection{Migration processes and urn models: a reminder}

The definition of a migration process appeared first in the
probabilistic literature (see e.g.~\cite{kelly}).
For the time being, we restrict the discussion to the one-dimensional case.
Consider a system of $N$ particles distributed among~$M$ sites,
with periodic boundary conditions.
Let $N_m$ be the number of particles at site $m=1,\dots,M$.
A migration process (or dynamical urn model) is defined by the following
Markovian dynamics:
(i) a departure (source) site $d$ is chosen at random,
uniformly among the $M$ sites of the system;
(ii) an arrival (target) site $a$ is chosen among the neighbors of $d$.
To be specific, the right neighbor ($a=d+1$) is chosen with probability $p$,
whereas the left neighbor ($a=d-1$) is chosen with
the complementary probability $q=1-p$;
(iii) a particle is transferred from site $d$ to site $a$
at a rate $W_{k,l}$ which only depends on the occupations
$k=N_d$ and $l=N_a$ of the two sites involved.
Of course, one has
\beq
W_{0,l}=0,
\label{zcons}
\eeq
since no particle can be removed from an empty site.

A one-dimensional migration process
is therefore entirely defined by the bias $p$ and the rate $W_{k,l}$.
A natural question is the following:
{\it What are the conditions on the bias~$p$ and
the rate $W_{k,l}$ for a one-dimensional migration process
to have a stationary-state product measure?}
This question was first addressed in~\cite{cocozza}
(for a review, see~\cite{lux}).

In this context, a stationary-state product measure means
that the probability of any configuration of the system
in its stationary state has the form
\beq
P(N_1,\dots,N_M)=\frac{1}{Z_{M,N}}\;p_{N_1}\dots p_{N_M}
\;\delta(N_1+\cdots+N_M,N),
\label{pm}
\eeq
where the factors $p_k$ are arbitrary,
and the partition function $Z_{M,N}$ reads
\beq
Z_{M,N}=\sum_{N_1,\dots,N_M}p_{N_1}\dots p_{N_M}\;\delta(N_1+\cdots+N_M,N).
\eeq

The answer to the above question is as follows:

\begin{itemize}
\item
For symmetric dynamics, i.e., when $p=1/2$,
the stationary state has a product measure characterized by the factor $p_k$,
if the rate $W_{k,l}$ obeys the condition
\beq
p_{k+1}p_lW_{k+1,l}=p_kp_{l+1}W_{l+1,k}.
\label{m1}
\eeq
The resulting stationary state is an equilibrium state.
The relation~(\ref{m1}) expresses detailed balance
with respect to this equilibrium state.

\item
For asymmetric dynamics, i.e., when $p\neq1/2$,
the stationary-state has a product measure
if the rate $W_{k,l}$ obeys both~(\ref{m1}) and the following condition:
\beq
W_{k,l}-W_{k,0}=W_{l,k}-W_{l,0}.
\label{m2}
\eeq
The stationary state is a genuine nonequilibrium steady state.
The condition~(\ref{m1}) does not express detailed balance any longer,
albeit a weaker property~\cite{lux},
referred to as pairwise balance~\cite{pairw}.
\end{itemize}

The partition function $Z_{M,N}$ associated with the product measure~(\ref{pm})
can be rewritten, using an integral representation of
the Kronecker delta function, as
\beq
Z_{M,N}=\oint\frac{\d z}{2\pi\i z^{N+1}}\,P(z)^M,
\label{contour}
\eeq
where the generating series of the factors $p_k$ reads
\beq
P(z)=\sum_{k\ge0}p_kz^k.
\eeq
The product measure~(\ref{pm}) is therefore characterized by $M$, $N$,
and the factor $p_k$ or, equivalently, by the generating function $P(z)$.
For a homogeneous system in the thermodynamic limit,
where $M$ and $N$ are simultaneously large,
with a fixed density $\rho=N/M$ of particles per site,
the contour integral in~(\ref{contour}) can be evaluated
by the saddle-point method.
The saddle-point value $z$, which is to be identified with the fugacity
in the grand canonical ensemble,
is related to the density $\rho$ through the equation
\beq
\frac{z P'(z)}{P(z)}=\rho.
\label{col}
\eeq
The distribution $f_k=\prob\{N_1=k\}$
of the occupations of any given site of the system
can be derived by summing the probability~(\ref{pm}) over $N_2,\dots,N_M$.
We thus obtain
\beq
f_k=p_k\,\frac{Z_{M-1,N-k}}{Z_{M,N}}.
\eeq
In the thermodynamic limit, this expression simplifies to
\beq
f_k=\frac{p_kz^k}{P(z)}.
\label{fdef}
\eeq

\subsection{The example of the zero-range process}

In the present context, the ZRP appears as the special
case of a migration process where
the rate $W_{k,l}$ only depends on the occupation of the departure site:
\beq
W_{k,l}=u_k,
\label{zrpdef}
\eeq
with $u_0=0$, by virtue of~(\ref{zcons}).
The condition~(\ref{m2}) is then automatically satisfied,
irrespective of the bias $p$.
Equation~(\ref{m1}) yields the following relation between the rate~$u_k$
and the factor $p_k$:
\beq
p_k=u_{k+1}p_{k+1},
\label{urel}
\eeq
up to a multiplicative constant,
which we set equal to unity by an appropriate choice of time unit.
The corresponding factor $p_k$
can be expressed in terms of the rate $u_k$ as follows:
\beq
p_0=1,\qquad p_k=\frac{1}{u_1\dots u_k}\qquad(k\ge1).
\label{pzrp}
\eeq
Reciprocally, to a given stationary-state product measure
characterized by the factor~$p_k$, there corresponds a unique ZRP dynamics
(up to a choice of time unit), whose rate reads
\beq
u_k=\frac{p_{k-1}}{p_k}\qquad(k\ge1).
\label{udef}
\eeq

\subsection{Characterization of migration processes with stationary-state
product measure}

We now give an explicit characterization of the migration processes
which admit a stationary-state product measure.

For symmetric dynamics, and for a given factor $p_k$,
the most general form of the rate obeying~(\ref{m1}) reads
\beq
W_{k,l}=\frac{p_{k-1}}{p_k}\,S_{k-1,l}\qquad(k\ge1),
\label{wsym}
\eeq
where the ratio $p_{k-1}/p_k$ is nothing but the rate $u_k$
of the corresponding ZRP, given by~(\ref{udef}),
whereas~$S_{k,l}$ is a symmetric function of $k$ and~$l$:
\beq
S_{k,l}=S_{l,k}.
\eeq
Besides the factor~$p_k$, the rate $W_{k,l}$ depends on an arbitrary
symmetric function~$S_{k,l}$ of {\it two} indices.

For asymmetric dynamics, and for a given factor~$p_k$,
the most general solution of~(\ref{m1}) and~(\ref{m2}) is determined
by the one-dimensional array of rates $\alpha_k=W_{k,0}$~\cite{kls3}.
Note that $\alpha_k$ is the rate at which an empty site is refilled,
by receiving one particle from a non-empty neighboring site
containing $k\ge1$ particles, and that~(\ref{zcons}) implies $\alpha_0=0$.
The above property can be shown as follows.
Consider all the indices $k$ and $l$ for a fixed value of the sum $k+l=n$,
and introduce the quantities
\beq
A_k=p_kp_{n-k}W_{k,n-k}\qquad(k=0,\dots,n).
\eeq
Equations~(\ref{m1}) and~(\ref{m2}) respectively become
\beq
A_k=A_{n+1-k},\qquad A_k-p_kp_{n-k}\alpha_k=A_{n-k}-p_kp_{n-k}\alpha_{n-k}.
\eeq
Combining these two equations yields
\beq
A_{k+1}-A_k=p_kp_{n-k}(\alpha_{n-k}-\alpha_k).
\eeq
The solution of this inhomogeneous difference equation
with initial value $A_0=0$ reads
\beq
A_k=\sum_{m=1}^kp_{k-m}p_{n-k+m}(\alpha_{n-k+m}-\alpha_{k-m}).
\eeq
We are thus left with the following expression for the rate $W_{k,l}$:
\beq
W_{k,l}=\frac{1}{p_kp_l}\sum_{m=1}^kp_{k-m}p_{l+m}(\alpha_{l+m}-\alpha_{k-m}).
\label{wasym}
\eeq
Besides the factor $p_k$,
the rate $W_{k,l}$ depends on an arbitrary function
$\alpha_k$ of {\it one} index.

\subsection{Definition of the target process}

We define the target process as the migration process where the rate
\beq
W_{k,l}=(1-\delta_{k,0})v_l
\label{tpdef}
\eeq
essentially depends on the occupation $l$ of the arrival (target) site.
Note the dual character of this definition with respect to the definition of
the ZRP, in that the roles of the departure and arrival sites
have been interchanged.
There is, however, a key difference between the two models,
coming from the presence of the constraint~(\ref{zcons}).
The latter,
which explicitly enters~(\ref{tpdef}) through the factor $(1-\delta_{k,0})$,
implies that the rate of the target process actually also bears some
dependence on the occupation $k$ of the departure site,
as it is constrained to vanish if $k=0$.
For the ZRP the same constraint does not
change the fact that the rate~$u_k$ only depends on the departure site;
it just imposes $u_0=0$.

For symmetric dynamics,
the target process always has a stationary-state product measure,
for any choice of the rate $v_k$.
Equation~(\ref{m1}) yields the following relation between the rate $v_k$
and the factor $p_k$:
\beq
p_{k+1}=v_kp_k,
\label{vrel}
\eeq
up to a multiplicative constant,
which is again set to unity by an appropriate choice of time unit.
The factor $p_k$ of the stationary-state measure
can thus be expressed in terms of the rate $v_k$ as follows:
\beq
p_0=1,\qquad p_k=v_0\dots v_{k-1}\qquad(k\ge1).
\label{pvdef}
\eeq
Reciprocally, to a given stationary-state product measure,
characterized by the factor~$p_k$,
there corresponds a unique symmetric target process
(up to a choice of time unit), whose rate reads
\beq
v_k=\frac{p_{k+1}}{p_k}.
\label{vdef}
\eeq

Equations~(\ref{urel}) and~(\ref{vrel}) show that the ZRP with rate $u_k$
and the target process with rate $v_k$ have the same stationary-state
product measure, i.e., the same factor $p_k$, if the rates obey
\beq
v_k=\frac{1}{u_{k+1}},
\label{uvdual}
\eeq
up to a multiplicative constant.
A target process and a ZRP related by this condition
are hereafter named {\it dual} to each other.

It is interesting to consider a more general class of migration processes,
where the rate has the form
\beq
W_{k,l}=(1-\delta_{k,0})u_k\,v_l.
\label{wgen}
\eeq
For symmetric dynamics, this model again has a stationary-state product
measure for any choice of $u_k$ and $v_l$.
Equation~(\ref{m1}) indeed yields
\beq
u_{k+1}p_{k+1}=v_kp_k,
\label{genrel}
\eeq
up to a multiplicative constant.
The factor $p_k$
can thus be expressed in terms of the $u_k$ and $v_k$ as follows:
\beq
p_0=1,\qquad p_k=\frac{v_0\dots v_{k-1}}{u_1\dots u_k}\qquad(k\ge1).
\eeq
This expression shows that the factor $p_k$
only depends on the ratio $v_k/u_{k+1}$.
This class of processes interpolates between the ZRP and the target process,
which are respectively recovered as the special cases where $v_l=1$ and $u_k=1$.
We finally notice that the rate~(\ref{tpdef}) of the target process,
and more generally the rate~(\ref{wgen}), is of the form~(\ref{wsym}),~with
\beq
S_{k,l}=v_kv_l.
\eeq

For asymmetric dynamics, the condition~(\ref{m2}) is very stringent.
The target process has a stationary-state product measure
if and only if the rate $v_k$ only assumes two values,
according to whether $k$ is zero or~not:
\beq
v_k=\left\{\matrix{v_0\hfill&(k=0),\cr v\hfill&(k\ge1).}\right.
\label{two}
\eeq
This also holds for the more general process defined by the rate~(\ref{wgen}).

For a generic asymmetric target process,
where the rate $v_k$ is not of the form~(\ref{two}),
the stationary-state measure is not a product measure.
It is not known explicitly, and can be expected to be a non-trivial
correlated measure in general.

\subsection{The `canonical target process' considered in this work}

The condensation phenomenon in the ZRP is usually investigated
using the rate~\cite{evans1,wis1,cg,gross}:
\beq
u_k=1+\frac{b}{k}\qquad(k\ge1),
\label{ucan}
\eeq
where the control parameter $b$ is a measure of the strength of interactions.
The minimality and exemplarity of this choice of rate suggest to call this
model the `canonical ZRP for condensation', or `canonical ZRP', for short.

Throughout the following we focus our attention onto
the target process dual to the canonical ZRP.
Its rate $v_k$ is therefore related to the rate~(\ref{ucan})
by the duality relation~(\ref{uvdual}).
We thus obtain
\beq
v_k=\frac{k+1}{k+b+1}\qquad(k\ge0).
\label{vcan}
\eeq
We name this process the `canonical target process'.

For symmetric dynamics,
it has already been shown above that the target process has the same
stationary-state product measure as the ZRP.
It therefore exhibits the same condensation transition.
We now give a brief reminder of the properties of this stationary-state measure.

In the absence of interactions ($b=0$),
the rate reads $u_k=v_k=1$, irrespective of $k$.
We have therefore $p_k=1$, so that $P(z)=1/(1-z)$.
The fugacity $z$ and the density $\rho$ are related by
\beq
z=\frac{\rho}{\rho+1},\qquad\rho=\frac{z}{1-z}.
\eeq
The distribution of the occupations~(\ref{fdef}) is a geometric distribution:
\beq
f_k=(1-z)z^k.
\label{fexpo}
\eeq

In the situation where $b$ is positive,
the rate $u_k$ is a decreasing function of the occupation $k$,
so that particles hop less easily out of more occupied sites.
Accordingly, the rate $v_k$ is an increasing function of $k$,
hence particles hop preferentially towards more occupied sites.
The rate~(\ref{ucan}) or~(\ref{vcan}) therefore corresponds to attractive
interactions between particles.
The model exhibits a trend toward segregation,
which leads to a thermodynamical condensation transition
if $b$ is strong enough.
It turns out that many characteristics of this condensation transition
are universal~\cite{loan,evans1,wis1,cg,gross,evans2,glcond}:
they only depend on the asymptotic behavior of the rate
$u_k$ (or $v_k$) at large $k$, i.e., essentially on the value of $b$.

For the choice of rate~(\ref{ucan}) or~(\ref{vcan}),
the factor $p_k$ of the stationary-state product measure
(see~(\ref{pzrp}) or~(\ref{pvdef})) reads
\beq
p_k=\frac{\Gamma(b+1)\,k!}{\Gamma(k+b+1)}=b\int_0^1(1-u)^{b-1}u^k\,\d u,
\label{pcan}
\eeq
so that
\beq
P(z)=b\int_0^1\frac{(1-u)^{b-1}}{1-zu}\,\d u.
\eeq
The factor $p_k$ falls off as a power law with exponent $b$ at large $k$:
\beq
p_k\approx\frac{\Gamma(b+1)}{k^b}.
\eeq
The canonical ZRP has a condensation transition
in the thermodynamic limit whenever the first moment of the factor $p_k$,
\beq
\rho_c=\sum_{k\ge1}k\,p_k=\frac{P'(1)}{P(1)},
\eeq
is convergent.
We have $P(1)=b/(b-1)$ and $P'(1)=b/((b-1)(b-2))$,
so that the critical density $\rho_c$ is finite for $b>2$, and reads
\beq
\rho_c=\frac{P'(1)}{P(1)}=\frac{1}{b-2}.
\eeq
This critical density separates a fluid phase and a condensed phase:

\begin{itemize}
\item
At the {\it critical density} ($\rho=\rho_c$, $z=1$),
the occupation distribution reads
\beq
f_k=\frac{p_k}{P(1)}=\frac{b-1}{b}\,p_k
\label{fkc}
\eeq
(see~(\ref{fdef})).
In particular the fraction of empty sites is
\beq
f_0=\frac{b-1}{b},
\label{f0c}
\eeq
whereas the distribution falls off as a power law for large occupations:
\beq
f_k\approx\frac{(b-1)\Gamma(b)}{k^b}.
\label{fktail}
\eeq
The statics and the dynamics of the model
exhibit many features of critical phenomena, including scaling and universality.

\item
In the {\it fluid phase} ($\rho<\rho_c$, $z<1$),
the occupation distribution $f_k$ falls off exponentially.

\item
In the {\it condensed phase} $(\rho>\rho_c)$, for a large finite system,
the particles are arranged so as to form a uniform critical background
and a macroscopic condensate, typically occupying one single site
and consisting of $N-M\rho_c=M(\rho-\rho_c)$ excess particles.
\end{itemize}

The stationary-state measure of the asymmetric canonical target process is not
a product measure in the presence of interactions,
i.e., for any non-zero value of the parameter~$b$,
because the rate~(\ref{vcan}) is not of the form~(\ref{two}).
This absence of a product measure also holds for
the asymmetric target process on higher-dimensional lattices.
The derivation of the conditions~(\ref{m1}),~(\ref{m2})
given in~\cite{lux} could indeed easily be extended to the case
of a biased one-particle dynamics on any higher-dimensional lattice.
The rest of this paper is devoted to a detailed investigation of this model.
The study of the one-dimensional totally asymmetric model
is presented in Sections~\ref{1D1} and~\ref{1D2},
whereas Section~\ref{2D} is devoted to a maximally asymmetric form of the model
on the two-dimensional square lattice.

\section{One-dimensional target process: theoretical analysis}
\label{1D1}

In this section, we consider the target process defined by the rate~(\ref{vcan})
in the totally asymmetric one-dimensional case ($p=1$).
The situation of most physical interest will turn out to be the regime
of a high density.

We begin the analysis of the model by exploring the consequences of
the existence of a conserved current.
Consider a large finite system, with periodic boundary conditions.
In the stationary state,
the mean current of particles through the system is conserved:
it assumes the same value $J$ through every bond.
The current through the bond between sites $m$ and $m+1$
is given by the mean value of the corresponding rate:
\beq
J=\mean{W_{N_m,N_{m+1}}}.
\label{jdef}
\eeq
The existence of this conserved current is expected to ensure
some degree of homogeneity of the stationary state.

To illustrate the method, let us first consider the totally asymmetric ZRP,
with rate~(\ref{ucan}).
The current therefore reads
\beq
J=\mean{u_{N_m}}.
\label{jzrpdef}
\eeq
Consider a typical configuration in the condensed phase ($\rho>\rho_c$).
The site where the condensate is located contains a macroscopic number of
particles, so that~(\ref{jzrpdef}) yields $J=1$
for the bond to the right of the condensate,
up to a negligible finite-size correction of order~$1/M$.
For all the other bonds, the relations~(\ref{fdef}) and~(\ref{urel}) yield
\beq
J=\sum_{k\ge1}\underbrace{u_k\,f_k}_{\ds zf_{k-1}}=z,
\eeq
where $z$ is the fugacity.
Equating the two above expressions for the current,
we recover well-known results for the totally asymmetric ZRP,
i.e., $z=z_c=1$ and $J=1$ throughout the condensed phase.

Let us now turn to the totally asymmetric target process,
with rate~(\ref{vcan}).
Equation~(\ref{jdef}) for the current now reads
\beq
J=\mean{(1-\delta_{N_m,0})v_{N_{m+1}}}.
\label{jtpdef}
\eeq
At variance with~(\ref{jzrpdef}),
this expression involves the occupations of two consecutive sites.

The high-density regime turns out to be the situation of most physical interest.
In this regime, at least some of the sites must have large occupations.
We are therefore led to distinguish between two types of sites:
\begin{itemize}
\item
B-sites (B for {\it big}), whose occupation is large, of the order of $\rho$.
\item
S-sites (S for {\it small}), whose occupation is small and fluctuating.
\end{itemize}

This distinction will be kept at a heuristic level throughout the following.
A typical high-density configuration therefore consists of
four types of bonds: BB, BS, SB, and~SS.
The typical values of the current in each type of bond
obey the following inequalities in the high-density limit:
\beq
J_\SS<\matrix{J_\SB\cr J_\BS}<J_\BB.
\label{jineq}
\eeq
Each factor in~(\ref{jtpdef}) is indeed less than unity,
and approaches unity in the limit where the involved occupation goes to
infinity.
In the high-density limit we have therefore $J_\BB=1$,
whereas $J_\BS=\mean{v_{N_m}}$ and $J_\SB=\mean{1-\delta_{N_m,0}}$,
where $m$ is a typical S-site, and~$J_\SS$ is smaller
than the last two expressions.

\begin{figure}[!tb]
\begin{center}
\includegraphics[angle=90,height=4truecm]{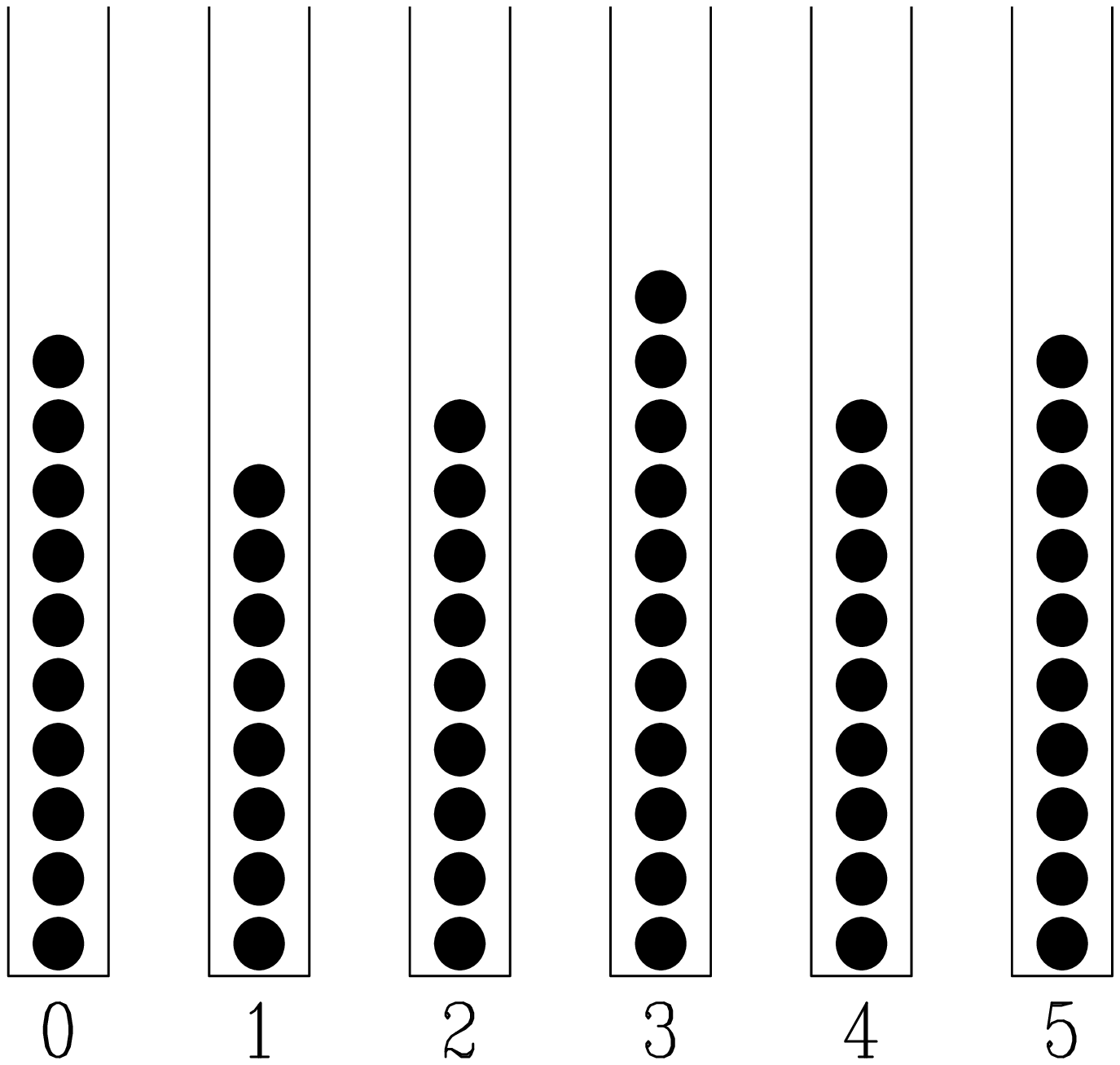}
{\hskip 60pt}
\includegraphics[angle=90,height=4truecm]{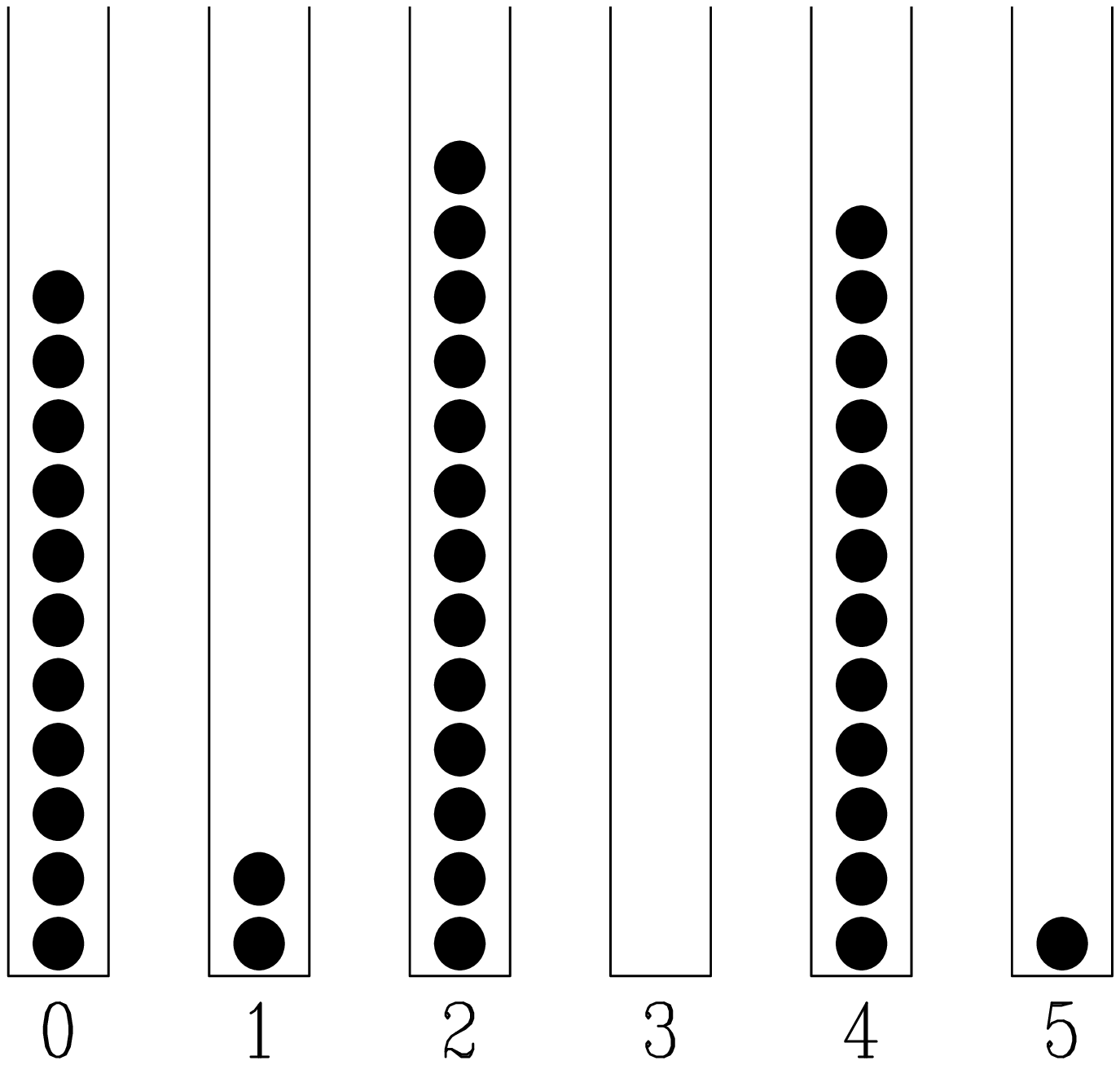}
\caption{\small
Typical configurations in the two possible scenarios
for the stationary state of the asymmetric one-dimensional target process
at high density.
Left: uniform configuration.
Right: alternating configuration.}
\label{figunialt}
\end{center}
\end{figure}

Thanks to the inequalities~(\ref{jineq})
the presence of a stable isolated condensate is excluded.
This would indeed correspond to a current profile where the two bonds
on either side of the condensate carry currents $J=J_\SB$ and $J=J_\BS$
which are significantly higher than the background current
$J=J_\SS$ of all the other bonds.
This non-uniform current distribution with a point defect would have the effect
that the condensate would soon dissolve into the background.
The inequalities~(\ref{jineq}) actually only leave out two scenarios
for typical stationary-state configurations in the high-density limit.
Both scenarios, described below and illustrated in Figure~\ref{figunialt},
correspond to spatially homogeneous phases.
We are therefore led to predict that the asymmetric target process
has no condensation transition at any finite density in one dimension.

\begin{itemize}
\item {\it Uniform scenario.}
In the uniform scenario, shown in the left panel of Figure~\ref{figunialt},
typical configurations are entirely made of B-sites,
whose typical occupation is around the density $\rho$.
These configurations carry a current $J=J_\BB$, i.e.,
\beq
J=1
\eeq
in the high-density limit, up to a correction of order $1/\rho$.

\item {\it Alternating scenario.}
In the alternating scenario, shown in the right panel of Figure~\ref{figunialt},
typical configurations are alternating, i.e., they have the form BSBSBSBS...

The asymptotic value of the current $J=J_\BS=J_\SB$
through a perfect alternating structure
in the high-density limit can be evaluated as follows.
Most particles belong to B-sites, whose typical occupation is very high,
around $2\rho$.
Every B-site therefore acts as a reservoir,
so that the occupations of the S-sites evolve independently from each other.
The master equations for the occupation distribution $f_k(t)$
of any given S-site read
\beqa
\frad{\d f_k}{\d t}&=&f_{k+1}+v_{k-1}f_{k-1}-(1+v_k)f_k\qquad(k\ge1),
\nonumber\\
\frad{\d f_0}{\d t}&=&f_1-v_0f_0.
\eeqa
The stationary-state solution of these equations is such that
$f_{k+1}=v_kf_k$.
It is therefore proportional to the factor $p_k$.
The properly normalized solution is given by~(\ref{fkc}).
To sum up, the occupations of the S-sites are independent variables,
whose distribution coincides with the critical occupation distribution
of the dual ZRP or, equivalently, of the symmetric target process.
We have
\beq
J=\sum_{k\ge0}\underbrace{v_k\,f_k}_{\ds f_{k+1}}=1-f_0.
\eeq
Using (\ref{f0c}), the current in the high-density limit reads
\beq
J=\frac{1}{b}.
\label{jcrit}
\eeq
This alternating scenario holds a priori whenever $b>1$.
The latter condition corresponds to $P(1)$ being finite,
so that~(\ref{fkc}) is a properly normalized probability distribution.
\end{itemize}

In order to have a hint on which of the two above scenarios is preferred,
it is interesting to first consider the simple example of a system of two sites.
In this case, there is only one kind of move for the particles,
namely from one site to the other, and only one degree of freedom,
the occupation $N_1=k$ of site number~1.
Indeed the occupation of site number~2 reads $N_2=N-k$.
The stationary-state occupation distribution~$f_k$ is
clearly equal to the equilibrium product measure
\beq
f_k=\frac{p_kp_{N-k}}{Z_{2,N}}\approx\frac{C_N}{k^b(N-k)^b}.
\eeq
When the number $N$ of particles is large,
the above distribution exhibits a segregation phenomenon
for any positive value of $b$ (see~\cite{glcond} for a more detailed analysis,
including an asymptotic analysis of the amplitude $C_N$).
The most probable configurations are those where almost all the
particles are at one site, i.e., either $k\ll N$ or $N-k\ll N$.
This simple example confirms that the target model, just as the ZRP,
manifests a trend toward segregation at high density.
It therefore suggests that the preferred scenario
is that of an alternating structure.

This picture can be corroborated and made more quantitative
by means of the following dynamical stability analysis of the uniform situation.
For the sake of generality, in this part of the analysis
we deal with the partially asymmetric target process with bias $p$.
Consider a configuration of the uniform scenario.
All the sites have very high local densities $\mean{N_m}=\rho_m(t)$.
These local densities obey the exact rate equation
\beqa
\frac{\d\rho_m}{\d t}
&=&p\mean{(1-\delta_{N_{m-1},0})v_{N_m}}
+q\mean{(1-\delta_{N_{m+1},0})v_{N_m}}\nonumber\\
&-&p\mean{(1-\delta_{N_m,0})v_{N_{m+1}}}
-q\mean{(1-\delta_{N_m,0})v_{N_{m-1}}}.
\label{rhorat}
\eeqa
In the high-density regime it is legitimate
to simplify the above equation in several respects.
The probability that a site is empty is negligible,
whereas $\mean{v_{N_m}}\approx v_{\mean{N_m}}\approx1-b/\mean{N_m}$.
We are thus left with
\beq
\frac{\d\rho_m}{\d t}\approx b
\left(\frac{p}{\rho_{m+1}}+\frac{q}{\rho_{m-1}}-\frac{1}{\rho_m}\right).
\eeq
Let us furthermore assume that the density profile is close
to being constant, i.e.,
\beq
\rho_m=\rho+\dr_m,
\eeq
with $\dr_m\ll\rho$.
The rate equation~(\ref{rhorat}) can then be linearized as
\beq
\frac{\d\,\dr_m}{\d t}\approx\frac{b}{\rho^2}
\left(\dr_m-p\,\dr_{m+1}-q\,\dr_{m-1}\right).
\eeq
The component of the density profile at wavevector $K$
therefore grows exponentially in time as $\hdr(K)\sim\exp(\sigma(K)\,t)$,
where the characteristic rate $\sigma(K)$ is given by the dispersion relation
\beq
\sigma(K)=\frac{b}{\rho^2}\left(1-p\,\e^{-\i K}-q\,\e^{\i K}\right).
\eeq
The growth rate $s(K)$ is given by the real part of $\sigma(K)$, which reads
\beq
s(K)=\frac{2b}{\rho^2}\;\sin^2\frac{K}{2},
\label{grow}
\eeq
irrespective of the bias $p$.
The expression~(\ref{grow}) for the growth rate of fluctuations
around the uniform situation is manifestly positive
for all values of the wavevector $K$.

The uniform scenario is therefore fully linearly unstable.
As a consequence, the alternating scenario is the preferred one.
Furthermore, the alternating structure is already appearing
as the most favored one within the stability analysis.
The most unstable mode indeed corresponds to $K=\pi$,
i.e., an alternating density modulation of the form $\dr_m\sim(-1)^m$.
A uniform high-density initial configuration
is therefore expected to smoothly relax to an alternating one
by the dynamics of the asymmetric target process.
This observation deserves, however, to be complemented with
the following caveat.
Consider the symmetric target process, corresponding to $p=1/2$.
The stability analysis still leads to the expression~(\ref{grow}),
so that the alternating mode is still the most favored one.
On the other hand, the stationary state is known to be described
by a product measure, and especially to have a single condensate.
This suggests that the relaxation dynamics of a uniform
high-density initial configuration will exhibit two stages:
first, a rather fast relaxation to an intermediate structure with
alternating fluctuations,
then, a coarsening evolution of the system toward its true fate,
by the merging of the excess particles into fewer and fewer condensate
precursors.
This two-stage relaxation should also hold for the dual ZRP
with a uniform high-density initial configuration.
The dynamical stability analysis of the uniform situation
in the ZRP indeed yields exactly the same expression~(\ref{grow}),
again irrespective of the bias $p$.

To close up, we mention that the expression~(\ref{grow})
also yields some hints on the time scales involved in the model.
The characteristic time of the most unstable mode, $T_\loc=1/s(\pi)$,
gives an estimate of the local relaxation time of the alternating structure,
at the spatial scale of two consecutive sites.
On the other hand,
the long-distance behavior of the dynamics is {\it antidiffusive}.
We have indeed formally $s(K)\approx -DK^2$ as $K\to0$,
with a small negative diffusion coefficient $D=-b/(2\rho^2)$.
For a large but finite system made of $M$ sites,
with periodic boundary conditions,
the longest characteristic time, $T_\gl=1/s(2\pi/M)$,
gives an estimate of the global relaxation time of the structure as a whole.
In the high-density regime, the characteristic times thus defined scale as
\beq
T_\loc\approx\frac{\rho^2}{2b},\qquad T_\gl\approx\frac{\rho^2M^2}{2\pi^2b}.
\label{times}
\eeq
The predicted divergence of both characteristic times with density
provides an a posteriori confirmation that the high-density regime
is indeed the most interesting one.

\section{One-dimensional target process: numerical results}
\label{1D2}

In this section we complement our analysis
of the target process in the asymmetric one-dimensional case,
by means of numerical simulations and scaling arguments.
For definiteness we restrict the study to the totally asymmetric
situation ($p=1$).
Furthermore, we set once for all $b=4$.
We successively consider features of the transient dynamics,
of the stationary-state measure, and of the stationary-state dynamics.
The main focus is on the scaling behavior of quantities of interest
in the high-density regime.
Two types of initial conditions are considered:
\begin{itemize}
\item Deterministic initial condition:
the occupations of all the sites are set equal to $N_m=\rho$
(provided the density $\rho$ is an integer).
\item Random initial condition:
the occupations $N_m$ are drawn independently at random
from the geometric distribution~(\ref{fexpo}) at density $\rho$.
\end{itemize}

We start by investigating the early stage of the dynamics,
where relatively fast rearrangements of particles
bring the system to a locally stationary state.
This stage of the relaxation dynamics can be monitored
by means of a local probe at one site.
We choose the reduced second moment of the occupations:
\beq
K(t)=\frac{1}{\rho^2}\sum_{k\ge0}k^2f_k(t),
\label{kdef}
\eeq
where $f_k(t)$ is the time-dependent distribution of the occupations.
The initial values of this quantity are
$K(0)=1$ for the deterministic initial condition,
and $K(0)=(2\rho+1)/\rho$ for the random one.

\begin{figure}[!tb]
\begin{center}
\includegraphics[angle=90,height=6truecm]{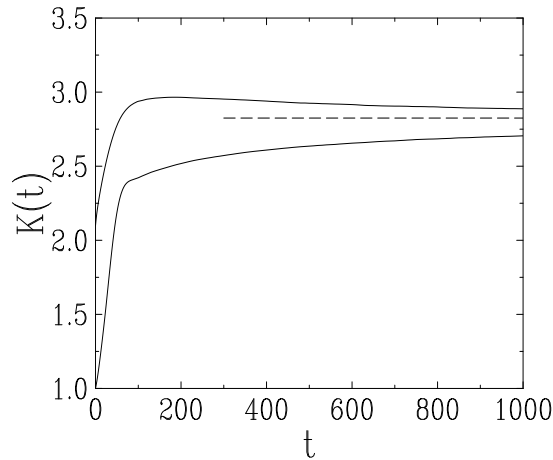}
\caption{\small
Plot of the reduced second moment $K(t)$ of the occupation distribution
in the one-dimensional fully asymmetric target process against time~$t$,
for $b=4$ and $\rho=10$.
Lower curve: deterministic initial condition.
Upper curve: random initial condition.
Horizontal dashed line: common limiting value $K\approx2.82$.}
\label{figkt}
\end{center}
\end{figure}

Figure~\ref{figkt} shows a plot of $K(t)$,
measured by means of a numerical simulation, against time $t$,
for $\rho=10$, and a deterministic and a random initial condition.
Each series of data are obtained by averaging over sufficiently many
histories in order to obtain a smooth signal
($10^4$ histories of a system of $10^4$ sites in this case).
The data exhibit a rather fast rise from their initial values,
and converge to a common limiting value, $K\approx2.82$,
extracted from data for much longer times, and shown as a dashed line.
The data for the deterministic initial condition (lower curve)
increase as a monotonic function of time and present a rather sharp shoulder,
whereas those for the random initial condition (upper curve)
exhibit a non-monotonic behavior with a very flat maximum.

\begin{figure}[!tb]
\begin{center}
\includegraphics[angle=90,height=6truecm]{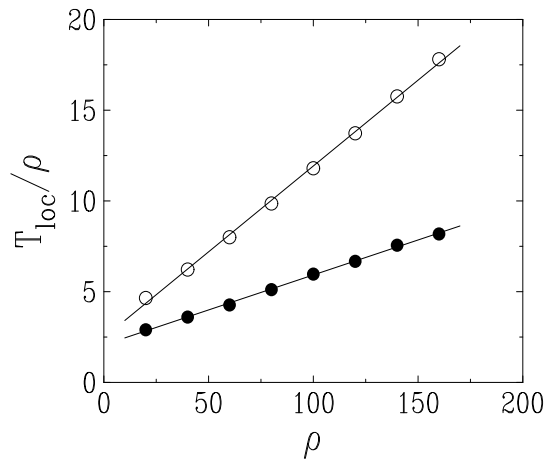}
\caption{\small
Plot of the local relaxation time $T_\loc$, divided by $\rho$,
against density~$\rho$.
Upper data (empty symbols): deterministic initial condition.
Lower data (full symbols): random initial condition.
Straight lines: least-squares fits with respective slopes 0.095 and 0.039.}
\label{figtloc}
\end{center}
\end{figure}

The characteristic time of the rise observed in $K(t)$
gives a measure of the local relaxation time.
More precisely, we define the local time~$T_\loc$
by the condition that~$K(t)$ is near the middle of its rise,
i.e., $K(T_\loc)=2$ for the deterministic initial condition,
and $K(T_\loc)=2.5$ for the random initial condition.
Figure~\ref{figtloc} shows a plot of the numerical values
of the local time so defined, divided by $\rho$, against $\rho$,
for both types of initial conditions.
Here and in subsequent figures,
statistical errors are comparable to the symbol size.
The least-squares fits of the two series of data suggest a growth of the form
\beq
T_\loc\approx A\rho^2+B\rho
\label{tlinquad}
\eeq
in the high-density regime of interest.
The amplitudes~$A$ and $B$ depend on the initial condition.
The least-squares fits shown on the plot
yield $A\approx0.095$ for a deterministic initial condition
and $A\approx0.039$ for a random initial condition.
These numbers, and especially the first one,
are comparable to the rough estimate coming from~(\ref{times}),
i.e., $A=1/(2b)=1/8=0.125$.

The initial condition is irrelevant
for what concerns later stages of the dynamics,
which correspond to the emergence of global features of the stationary state.
Hereafter we choose to work with a random initial condition.
Before we turn to an analysis of these late stages,
it is worth taking a glance at the spatial structure
of the stationary state in the high-density regime.

\begin{figure}[!tb]
\begin{center}
\includegraphics[angle=90,height=6truecm]{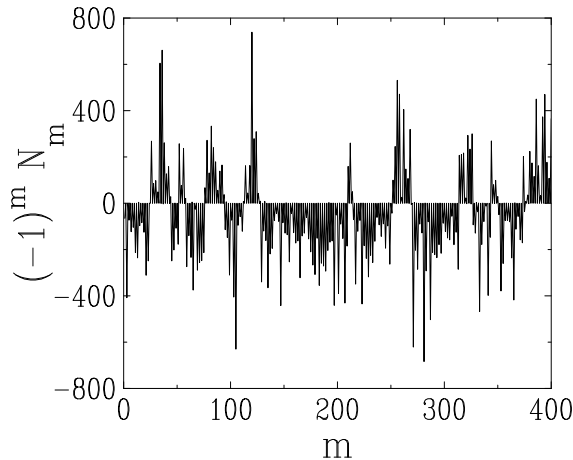}
\caption{\small
Plot of $(-1)^mN_m$ against the position $m$ of the site,
for a typical stationary-state configuration for $\rho=50$,
emphasizing the alternating structure.}
\label{figtrack}
\end{center}
\end{figure}

Figure~\ref{figtrack} shows a typical stationary
occupation profile for $\rho=50$.
In order to better reveal the alternating structure predicted
in Section~\ref{1D1},
we have plotted $(-1)^mN_m$ against the position $m$ of the site.
The alternating structure clearly emerges from this representation:
domains where the signal is positive (resp.~negative)
correspond to domains where the B-sites are the even (resp.~odd) sites.

In order to turn this observation into a quantitative measurement,
let us introduce the concept of {\it defects}.
Roughly speaking, a defect is a site around which the structure is not perfectly
of the alternating form BSBSBSBS...
and a domain is any stretch between two consecutive defects.
More precisely, the site~$m$ is considered as a defect whenever
its occupation $N_m$ is neither a maximum nor a minimum of the density profile.
Equivalently,~$N_m$ is between $N_{m-1}$ and $N_{m+1}$,
i.e., the product $(N_m-N_{m-1})(N_m-N_{m+1})$ is negative.
This definition pinpoints 24 defects
in the configuration shown in Figure~\ref{figtrack}.
This number is slightly above the number of domains visible with
the naked eye, i.e.,~18, because some of the domains are microscopic.
Most defects can be viewed either as a BB sequence or as an SS sequence.
We shall return later on to the dynamics of these defects.
The density of defects $R$, i.e., the mean number of defects
per unit length, reads
\beq
R=\bigmean{\Theta\Bigl(-(N_m-N_{m-1})(N_m-N_{m+1})\Bigr)},
\eeq
where the Heaviside step function on the integers is defined as
\beq
\Theta(n)=\left\{\matrix{
1\quad\hbox{for}\;\;n>0,\hfill\cr
0\quad\hbox{for}\;\;n\le0.}\right.
\eeq
The inverse of the density of defects,
\beq
\xi=\frac{1}{R},
\eeq
is interpreted as the mean size of a domain or, equivalently,
as the {\it coherence length} of the alternating structure.

\begin{figure}[!tb]
\begin{center}
\includegraphics[angle=90,height=6truecm]{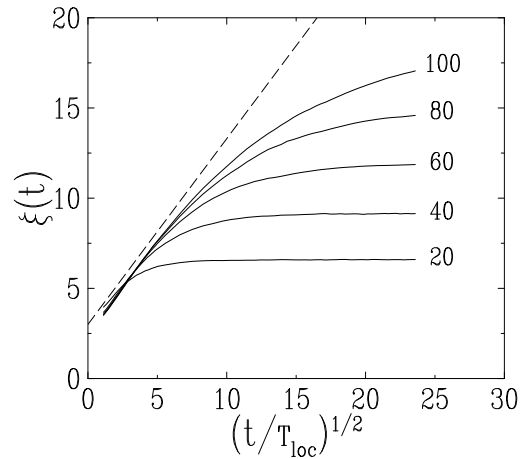}
\caption{\small
Plot of the mean domain size $\xi(t)$
against the reduced time $(t/T_\loc)^{1/2}$,
for various values of the density $\rho$, indicated on the curves.
The dashed straight line starting from $\xi(0)=3$
illustrates the coarsening law~(\ref{coa}).}
\label{figd}
\end{center}
\end{figure}

Let us now return to the late stages of the dynamics,
starting from a random initial condition.
In order to characterize the growth of the alternating structure,
we have measured the time dependence of the mean domain size $\xi(t)=1/R(t)$.
Figure~\ref{figd} shows a plot of~$\xi(t)$,
for various values of the density $\rho$.
For a random initial condition, we have $\xi(0)=3$.
Consider indeed the initial values of $N_{m-1}$, $N_m$, and $N_{m+1}$.
The probability that these three independent random numbers
obey either of the inequalities $N_{m-1}<N_m<N_{m+1}$
or $N_{m-1}>N_m>N_{m+1}$ is equal to 1/6 in the high-density regime
(neglecting the fact that these are integer variables,
which may coincide with a small but nonzero probability).
Hence $R(0)=2\times1/6=1/3$ and $\xi(0)=1/R(0)=3$.
The data are plotted against the reduced time variable $(t/T_\loc)^{1/2}$,
for each value of the density~$\rho$,
where $T_\loc$ is taken from the lower data of Figure~\ref{figtloc}.
The observed common initial linear behavior,
shown as a dashed straight line starting from the known value $\xi(0)=3$,
demonstrates that the mean domain size grows according to the coarsening~law
\beq
\xi(t)\sim(t/T_\loc)^{1/2},
\label{coa}
\eeq
before it saturates to a density-dependent
stationary-state value, simply denoted by $\xi$.
The duration of the coarsening process, before the stationary state is reached,
defines the global relaxation time $T_\gl$ of the problem.
By inverting the relation~(\ref{coa}),
we predict that the latter time grows as $T_\gl\sim T_\loc\,\xi^2$.
This relation between $T_\loc$ and $T_\gl$ is in agreement with~(\ref{times}),
where the stationary-state mean domain size $\xi$ plays the role
of the system size $M$.

We now investigate a few characteristic features
of the nonequilibrium stationary state of the asymmetric target process,
emphasizing their scaling behavior at high density.
We start with the mean domain size~$\xi$.
Its scaling behavior at high density can be predicted by the following argument.
Roughly speaking, a defect can be thought of as a B-site whose occupation
is accidentally as small as that of an S-site, i.e., finite.
Anticipating the scaling law~(\ref{bsca}),
and assuming a linear rise for the scaling function~$F(x)$,
we find that the probability of such an event scales as $1/\rho^2$.
This argument leads to an asymptotic
quadratic growth of the mean domain size of the~form
\beq
\xi\approx\Xi\,\rho^2.
\label{xiquad}
\eeq

\begin{figure}[!tb]
\begin{center}
\includegraphics[angle=90,height=6truecm]{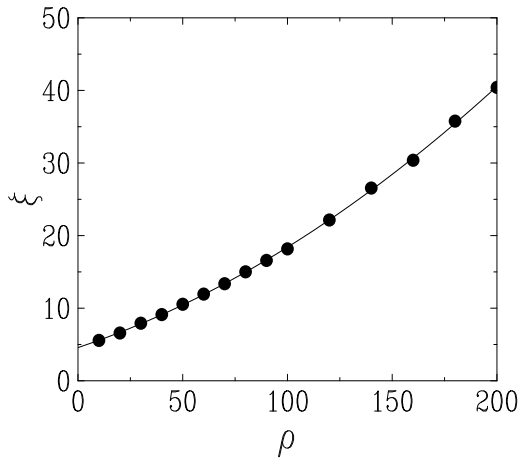}
\caption{\small
Plot of the mean domain size $\xi$ in the stationary state
against density~$\rho$.
Full line: second-degree polynomial fit to the data
with leading coefficient $\Xi=416\times10^{-6}$.}
\label{figxi}
\end{center}
\end{figure}

Figure~\ref{figxi} shows a plot of the stationary-state value
of $\xi$ against density $\rho$, for densities up to $\rho=200$.
The second-degree polynomial fit to the data is compatible with
our expectation~(\ref{xiquad}).
The numerical value for the prefactor, $\Xi\approx4\times10^{-4}$,
is however found to be very small.
Partly as a consequence of this smallness,
the data exhibit large corrections to the above asymptotic law
for values of the density accessible to numerical simulations.
At variance with the case of $T_\loc$, plotted in Figure~\ref{figtloc},
we found no way to unambiguously characterize these corrections.
The quadratic growth~(\ref{xiquad}) of~$\xi$
corresponds to a very fast growth of the global relaxation time:
\beq
T_\gl\sim T_\loc\,\xi^2\sim\rho^6.
\eeq

\begin{figure}[!tb]
\begin{center}
\includegraphics[angle=90,height=6truecm]{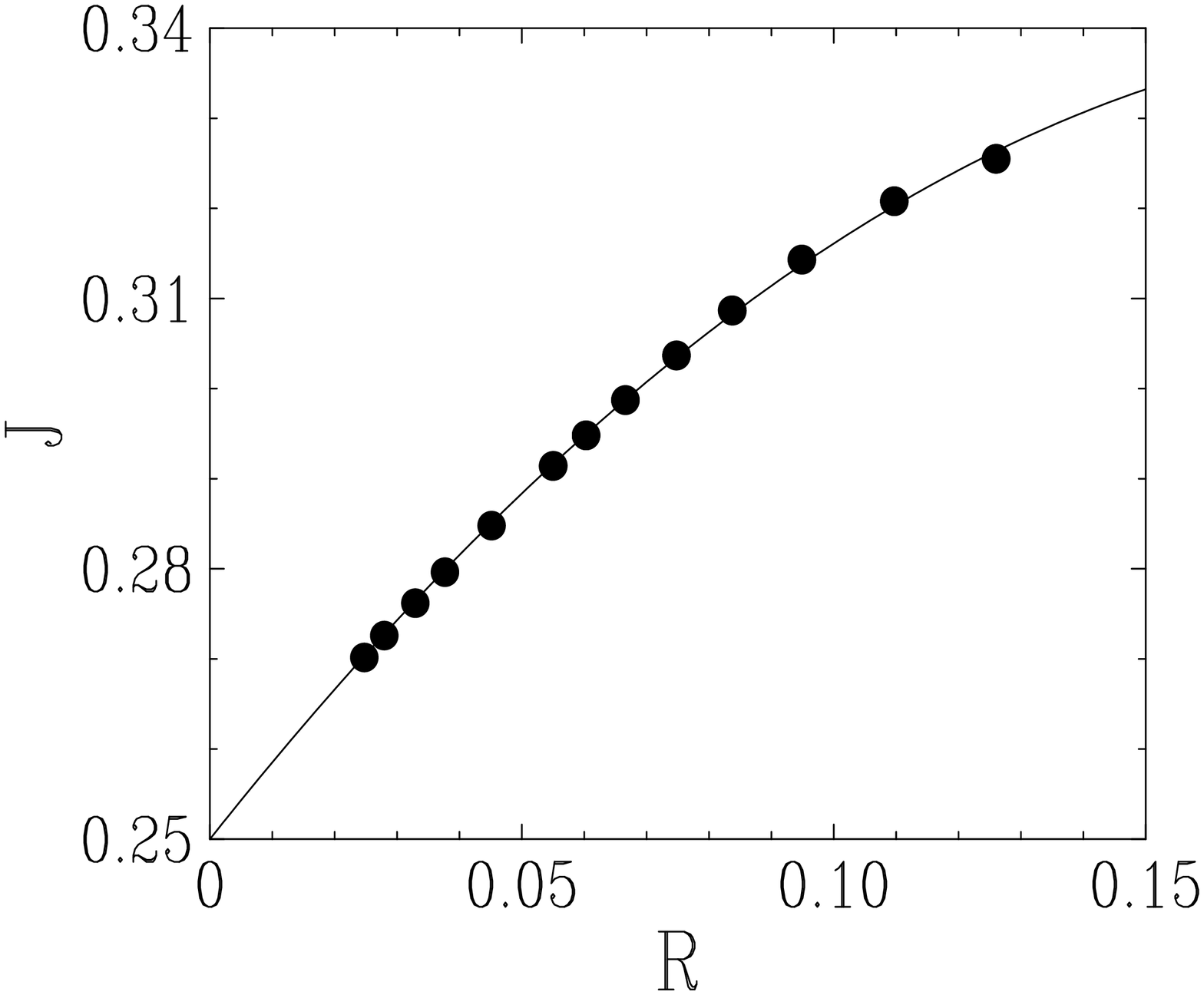}
{\hskip 15pt}
\includegraphics[angle=90,height=6truecm]{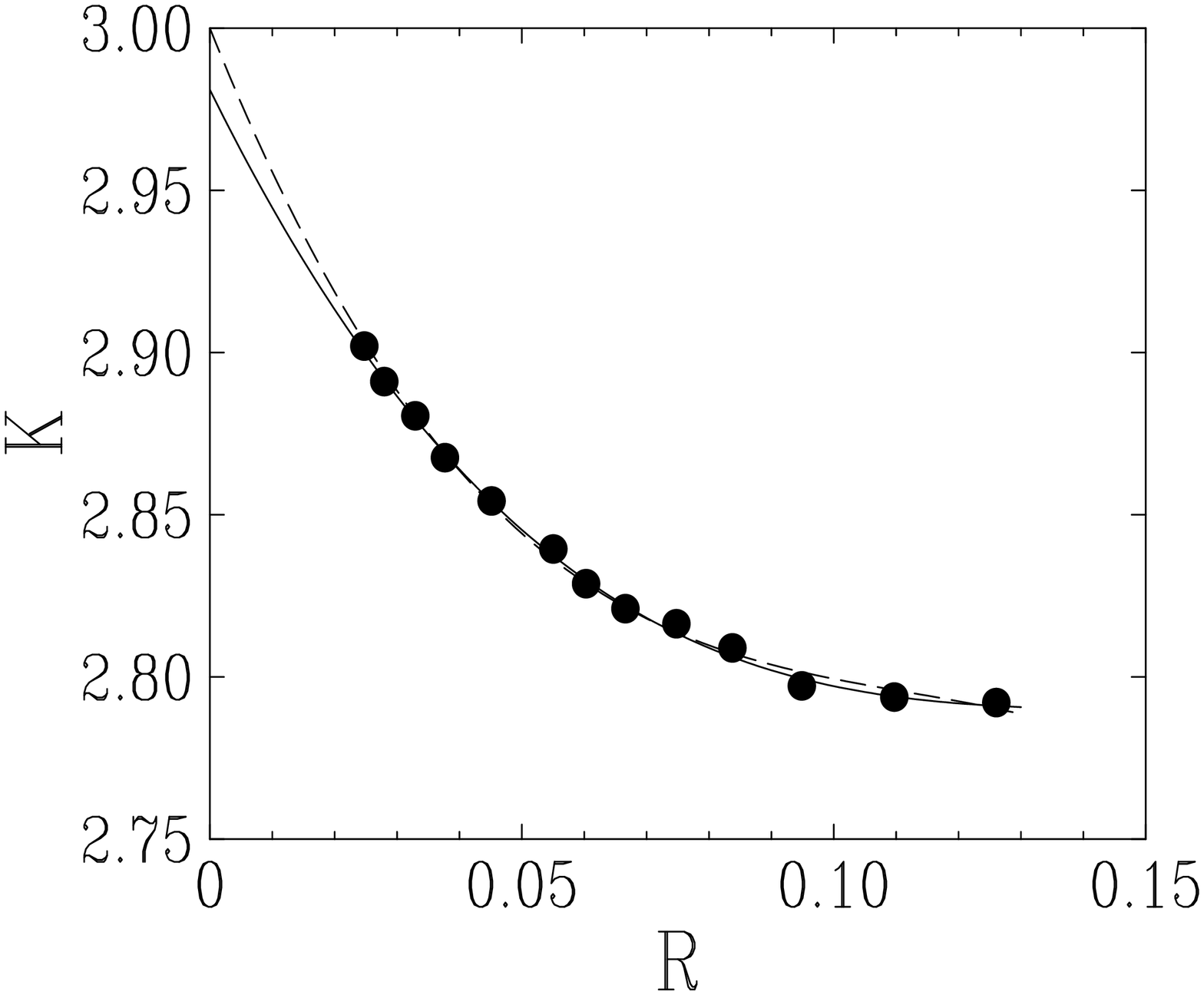}
\caption{\small
Plot of stationary values of the current $J$ (left) and of the
reduced second moment~$K$ (right),
against the stationary density of defects $R=1/\xi$.
Symbols: data for densities ranging from $\rho=30$ to 200.
Full lines: polynomial fits, of degree 2 for $J$ and 3 for $K$,
yielding the extrapolated values $J_\infty\approx0.250$
and $K_\infty\approx2.98$.
Dashed line: constrained polynomial fit of degree 3 for $K$,
imposing $K_\infty=3$ (see text).}
\label{figjk}
\end{center}
\end{figure}

Figure~\ref{figjk} shows a plot of the stationary-state values
of the current $J$ and of the reduced second moment $K$
of the occupation distribution,
against the stationary-state density of defects $R=1/\xi$.
The second-degree polynomial fit for $J$ yields $J_\infty\approx0.250$,
in excellent quantitative agreement with
the limiting value $J=1/b=1/4$ predicted in~(\ref{jcrit}).
A third-degree polynomial fit for $K$ yields the
limiting value $K_\infty\approx2.98$.
This number is very close to the value $K_\infty=3$ corresponding to the
trial scaling function $F_0(x)$ given in~(\ref{ftrial}).
This proximity suggests that the value $K_\infty=3$ could be exact.
Imposing the constraint $K_\infty=3$ indeed hardly alters
the quality of the fit (dashed line).

\begin{figure}[!tb]
\begin{center}
\includegraphics[angle=90,height=6truecm]{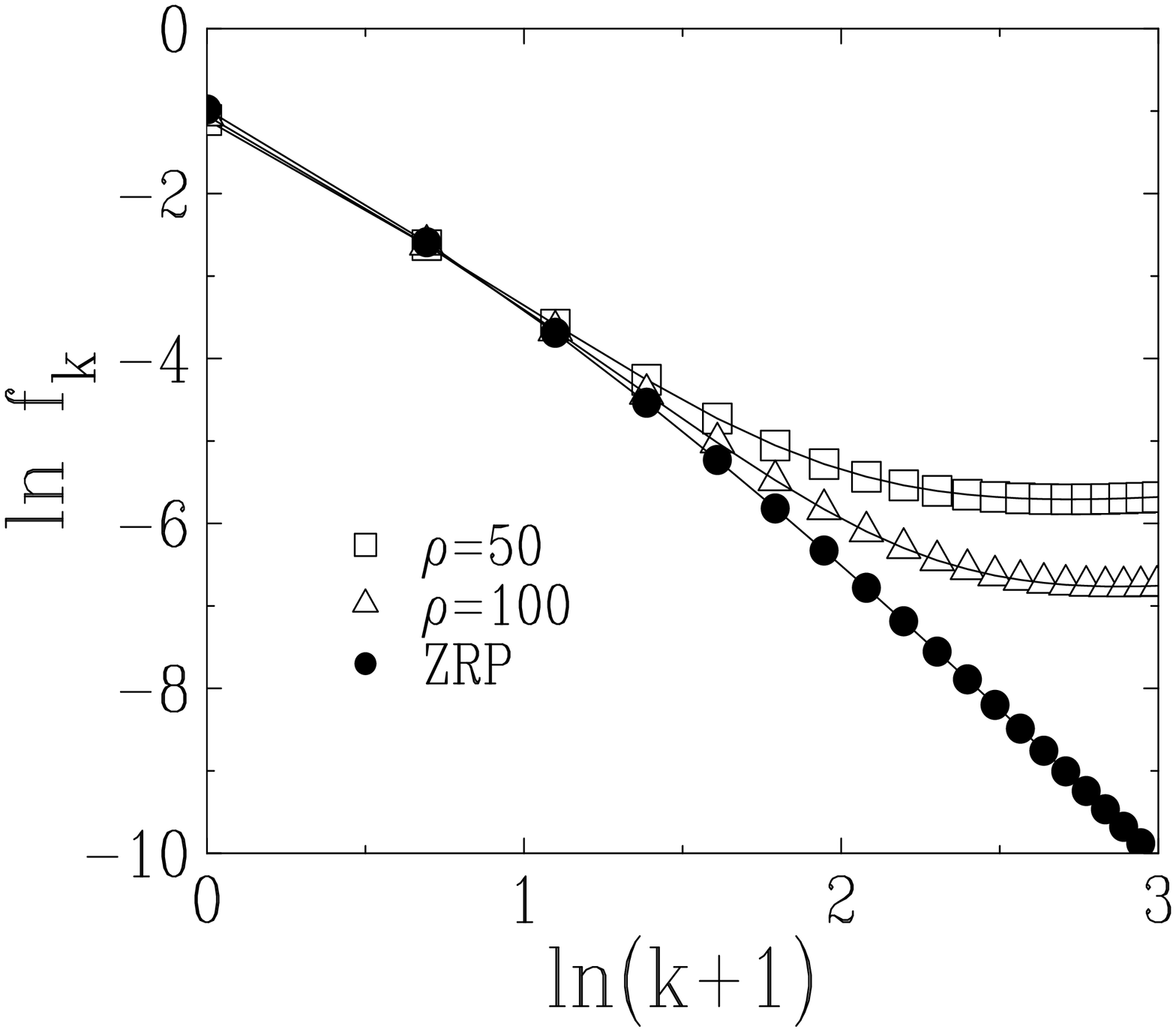}
{\hskip 15pt}
\includegraphics[angle=90,height=6truecm]{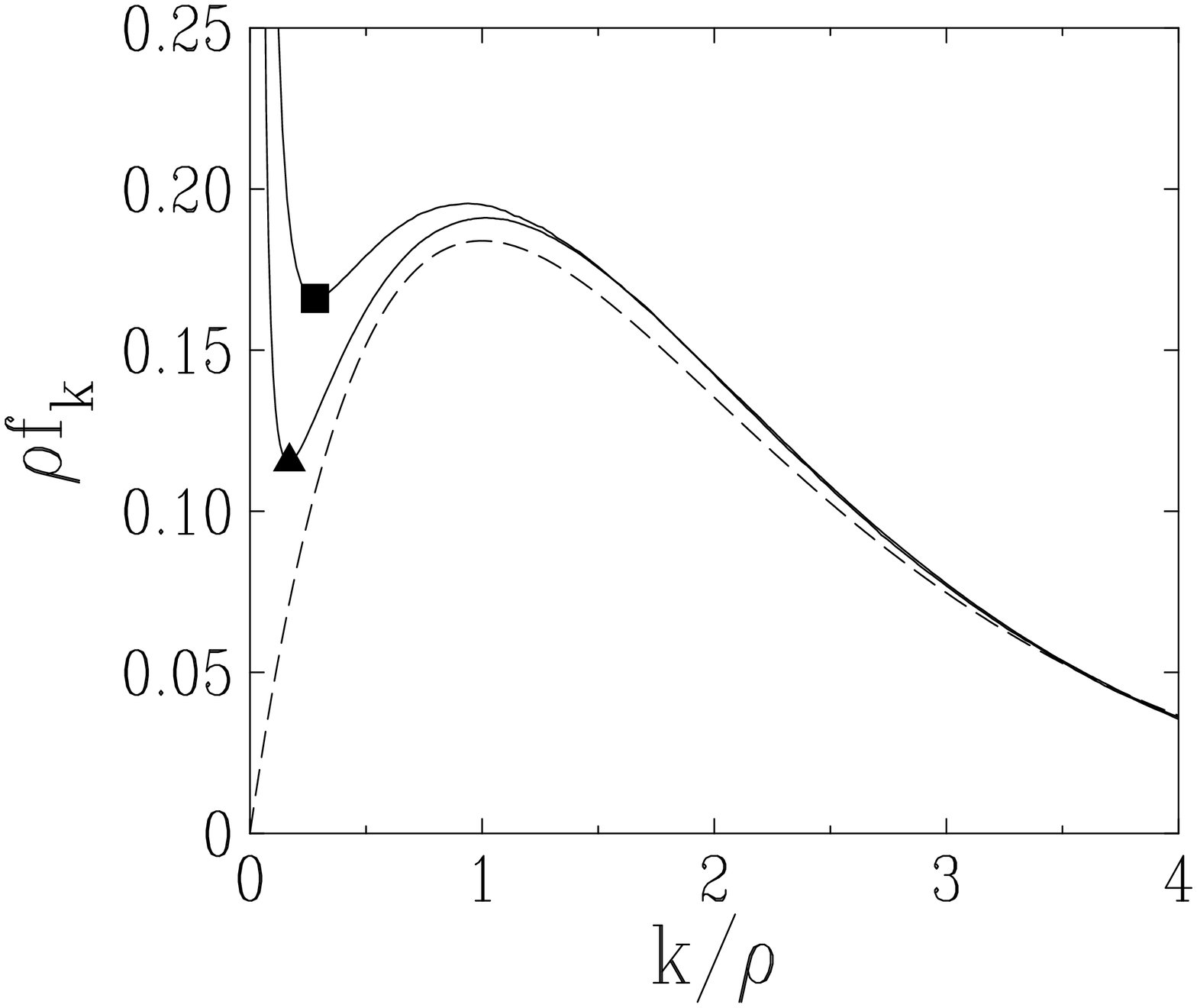}
\caption{\small
Stationary-state occupation distribution $f_k$.
The left panel shows data for $k\ll\rho$ (S-sites).
Empty symbols: plot of $\ln f_k$ against $\ln(k+1)$
for $\rho=50$ (upper data) and 100 (lower data).
Full symbols (labeled ZRP): plot of $\ln (f_k/2)$,
where $f_k$ is the critical distribution of the dual ZRP.
The right panel shows data for $k$ comparable to the density $\rho$ (B-sites).
Full lines: scaling plots of the product $\rho f_k$ against $k/\rho$
for $\rho=50$ (upper data) and 100 (lower data).
Symbols: minimum value of $f_k$.
Dashed line: trial scaling function $F_0(x)$ defined in~(\ref{ftrial}).}
\label{figf}
\end{center}
\end{figure}

In complete agreement with the alternating scenario
depicted in Section~\ref{1D1},
the occupation distribution $f_k$ in the stationary state
consists of two distinct components with equal weights,
which respectively describe S-sites and B-sites.
These components are emphasized in Figure~\ref{figf},
where the data for $f_k$ for the same values of the density, $\rho=50$ and 100,
are plotted in two different ways.
The left panel shows a plot of $\ln f_k$ against $\ln(k+1)$
for moderate values of the occupation $k$, up to~20.
These are, roughly speaking, the S-sites.
The data for the smaller values of~$k$
are very close to $1/2$ times the critical occupation distribution
of the dual ZRP or, equivalently, of the symmetric target process,
given by~(\ref{fkc}), also shown on the plot (full symbols).
The range of values of $k$ over which the agreement holds
is observed to get larger for larger densities.
This is a convincing confirmation of the prediction made
in Section~\ref{1D1} that the S-sites are critical in the stationary state.
The right panel shows a scaling plot of the product $\rho f_k$ against $k/\rho$,
for larger values of $k$, comparable to the density $\rho$.
These are, roughly speaking, the B-sites.
The data exhibit a scaling law of the form
\beq
f_k\approx\frac{1}{\rho}\;F\!\left(\frac{k}{\rho}\right).
\label{bsca}
\eeq
The scaling function $F(x)$ obeys the sum rules
\beq
\int_0^\infty F(x)\,\d x=1/2,\quad\int_0^\infty xF(x)\,\d x=1,\quad
\int_0^\infty x^2F(x)\,\d x=K_\infty.
\eeq
The first two equalities express that the fraction of B-sites is $1/2$
and that their mean occupation is $2\rho$, whereas
the third one is a rewriting of the definition of~$K_\infty$.
The scaling function is observed to be rather uniformly well approximated by
the trial scaling function
\beq
F_0(x)=\frac12\,x\,\e^{-x},
\label{ftrial}
\eeq
shown on the right panel of Figure~\ref{figf} as a dashed line.
The scaling function~(\ref{ftrial}) corresponds to $K_\infty=3$.
The closeness of this number to the extrapolated value $K_\infty\approx2.98$
opens up the possibility that the scaling function $F(x)$ is exactly given
by $F_0(x)$.
In any case, the linear rise of the trial function $F_0(x)$
seems to be shared by the true scaling function $F(x)$.

Both components of the occupation distribution shown in Figure~\ref{figf},
respectively corresponding to S-sites and B-sites,
are separated by a minimum in the occupation distribution.
In the high-density regime
this minimum takes place for a crossover occupation $k_\star$
such that the estimates $1/k^b$ (see~(\ref{fktail})) and $k/\rho^2$
(assuming a linear rise for the scaling function $F(x)$) are comparable.
We thus obtain
\beq
k_\star\sim\rho^{2/(b+1)},\qquad f_{k_\star}\sim\rho^{-2b/(b+1)}.
\label{xover}
\eeq
These estimates make sense as soon as the crossover occupation obeys
$k_\star\ll\rho$.
We thus recover the condition~$b>1$ for the validity
of the alternating scenario.
The mean occupation $\rho_S$ of the S-sites has the finite asymptotic value
$\rho_c$ for $b>2$,
whereas it scales as $\rho_S\sim k_\star^{2-b}\sim\rho^{2(2-b)/(b+1)}$
in the high-density regime for $1<b<2$.

We close up this section
with an investigation of the stationary-state dynamics of defects.
For a large but finite density $\rho$,
there is a small density of defects $R=1/\xi\sim1/\rho^2$ (see~(\ref{xiquad})).
These defects cannot stay immobile.

\begin{itemize}
\item
Consider indeed a BB defect, made of two consecutive B-sites.
The current between the two B-sites, $J_\BB=1$, exceeds the mean current
$J=1/b$ through the system.
As a consequence, particles flow from the left B-site of the defect
into the right one at a rate $\omega_\BB=J_\BB-J=(b-1)/b$.
After a time of the order of
\beq
\tau_\BB\approx\frac{2\rho}{\omega_\BB}=\frac{2b\rho}{b-1},
\eeq
the left B-site is emptied.
This is the first reaction of~(\ref{bbss}).
An SS defect is thus formed one site to the left of the original BB defect.

\item
Consider now an SS defect, made of two consecutive S-sites.
The current between the two S-sites, $J_\SS\approx1/b^2$,
is smaller than the mean current $J$.
As a consequence, particles flow from the B-site to the left of the defect
into the left S-site at a rate $\omega_\SS=J-J_\SS\approx(b-1)/b^2$.
The left S-site is thus soon (i.e., after a time which does not
grow proportionally to $\rho$) turned to a B-site.
This is the second reaction of~(\ref{bbss}).
A BB defect is thus formed one site to the left of the original~SS defect.
\end{itemize}

\begin{figure}[!tb]
\begin{center}
\includegraphics[angle=90,height=6truecm]{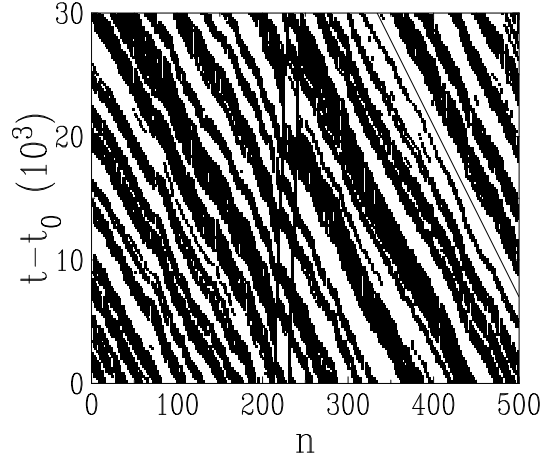}
\caption{\small
Space-time plot of the stationary domain pattern for $\rho=100$.
Black (resp.~white) areas show the regions of space-time
where the S-sites are the odd (resp.~the even) sites.
The slope of the straight line in the right part of the plot
yields $V\approx-0.0072$.}
\label{figplot}
\end{center}
\end{figure}

The discussion can be summarized in the form of the following reactions
\beq
\BB\to\SB,\qquad\SS\to\BS.
\label{bbss}
\eeq
The typical history of a single defect therefore looks as follows,
where time runs from bottom to top,
for the sake of consistency with Figure~\ref{figplot}:
\[
t\uparrow\quad\matrix{
\BS\BS\bfBB\SB\SB\SB\SB\SB\SB\SB\cr
\BS\BS{\rm B}\bfSS{\rm B}\SB\SB\SB\SB\SB\SB\cr
\BS\BS\BS\bfBB\SB\SB\SB\SB\SB\SB\cr
\BS\BS\BS{\rm B}\bfSS{\rm B}\SB\SB\SB\SB\SB\cr
\BS\BS\BS\BS\bfBB\SB\SB\SB\SB\SB\cr
\BS\BS\BS\BS{\rm B}\bfSS{\rm B}\SB\SB\SB\SB\cr
\BS\BS\BS\BS\BS\bfBB\SB\SB\SB\SB\cr
\BS\BS\BS\BS\BS{\rm B}\bfSS{\rm B}\SB\SB\SB\cr
n\to}
\]

The time it takes for a defect to move two sites to the left
is therefore equal to $\tau_\BB$ on average.
As a consequence, the pattern of defects and domains is
advected with an upstream (negative) velocity $V\approx-2/\tau_\BB$, i.e.,
\beq
V\approx-\frac{b-1}{b\rho}.
\label{vbal}
\eeq
This scenario is confirmed by Figure~\ref{figplot},
showing a space-time plot of the stationary dynamics
of the domain pattern for $\rho=100$.
Equation~(\ref{vbal}) predicts $V\approx-0.0075$,
in reasonably good agreement with the observed value $V\approx-0.0072$.
The plot shows that the whole advected domain pattern
behaves more or less as a rigid body over spatial scales
much larger than the mean domain size $\xi$.

\section{Two-dimensional target process}
\label{2D}

We now consider the canonical target process
defined by the rate~(\ref{vcan}), on the square lattice
with unit vectors $\bfe_1$, $\bfe_2$.
In order to have a genuine two-dimensional model
and to maximize the asymmetry, we choose the following rule:
particles hop either East (displacement $\bfe_1$)
or North (displacement $\bfe_2$) with equal probabilities.
In other words, if the departure site is $\bfd=(m,n)$,
the arrival site is chosen to be either $\bfa=(m+1,n)=\bfd+\bfe_1$,
or $\bfa=(m,n+1)=\bfd+\bfe_2$, with probability $1/2$.
The bias, i.e., the mean displacement proposed
to a particle, $\bfb=(\bfe_1+\bfe_2)/2$, is along the North-East direction.

\subsection{Heuristic argument}

In the one-dimensional situation, the existence of a conserved current
was instrumental in order to discriminate between possible scenarios
for the stationary state of the model.
In the present situation, however, the current $\J$ is a two-dimensional
vector.
The condition that~$\J$ be conserved in the stationary state
is less stringent than in the one-dimensional case.
On spatial scales much larger than the lattice spacing,
it is reasonable to use the continuum formalism.
Within this framework, the conservation law reads $\nabla\cdot\J=0$.
The symmetry of the dynamical rules implies that $\J$
is aligned with the bias $\bfb$.
Current lines are therefore parallel straight lines along this direction.
The conservation law implies that the magnitude of the current
is constant along each current line,
but may well vary in the transversal direction from one current line to another.
In particular the existence of an {\it extended condensate} is allowed,
in the form of a one-dimensional structure
elongated along the direction of the bias.
This opens up the possibility of having an unconventional
type of condensation transition.
Such a phenomenon is indeed observed in the numerical simulations described
hereafter.

\subsection{Numerical results}

We choose once for all the value $b=4$ in the numerical simulations
of the two-dimensional target process defined above.

\begin{figure}[!tb]
\begin{center}
\includegraphics[angle=90,height=6truecm]{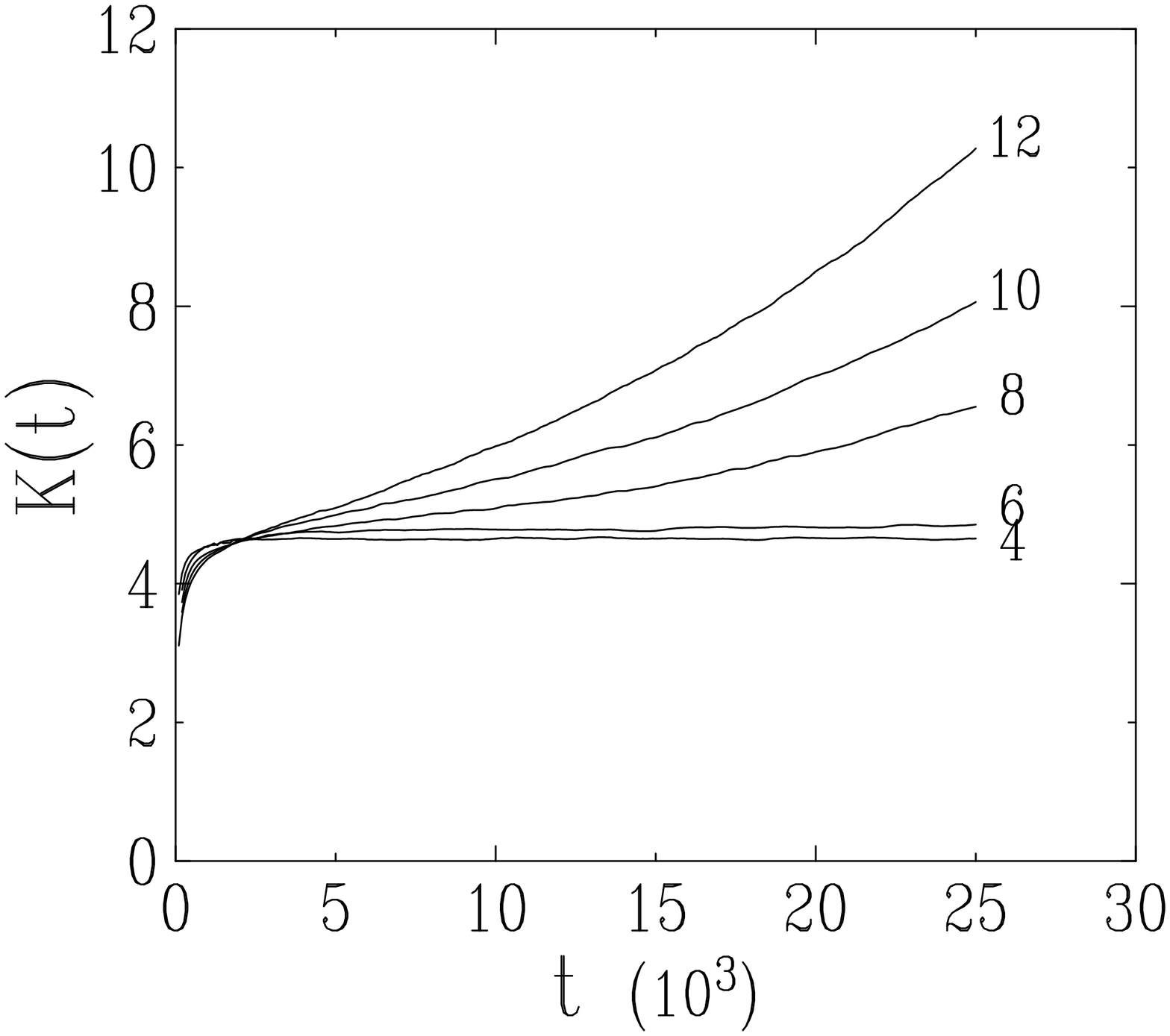}
{\hskip 15pt}
\includegraphics[angle=90,height=6truecm]{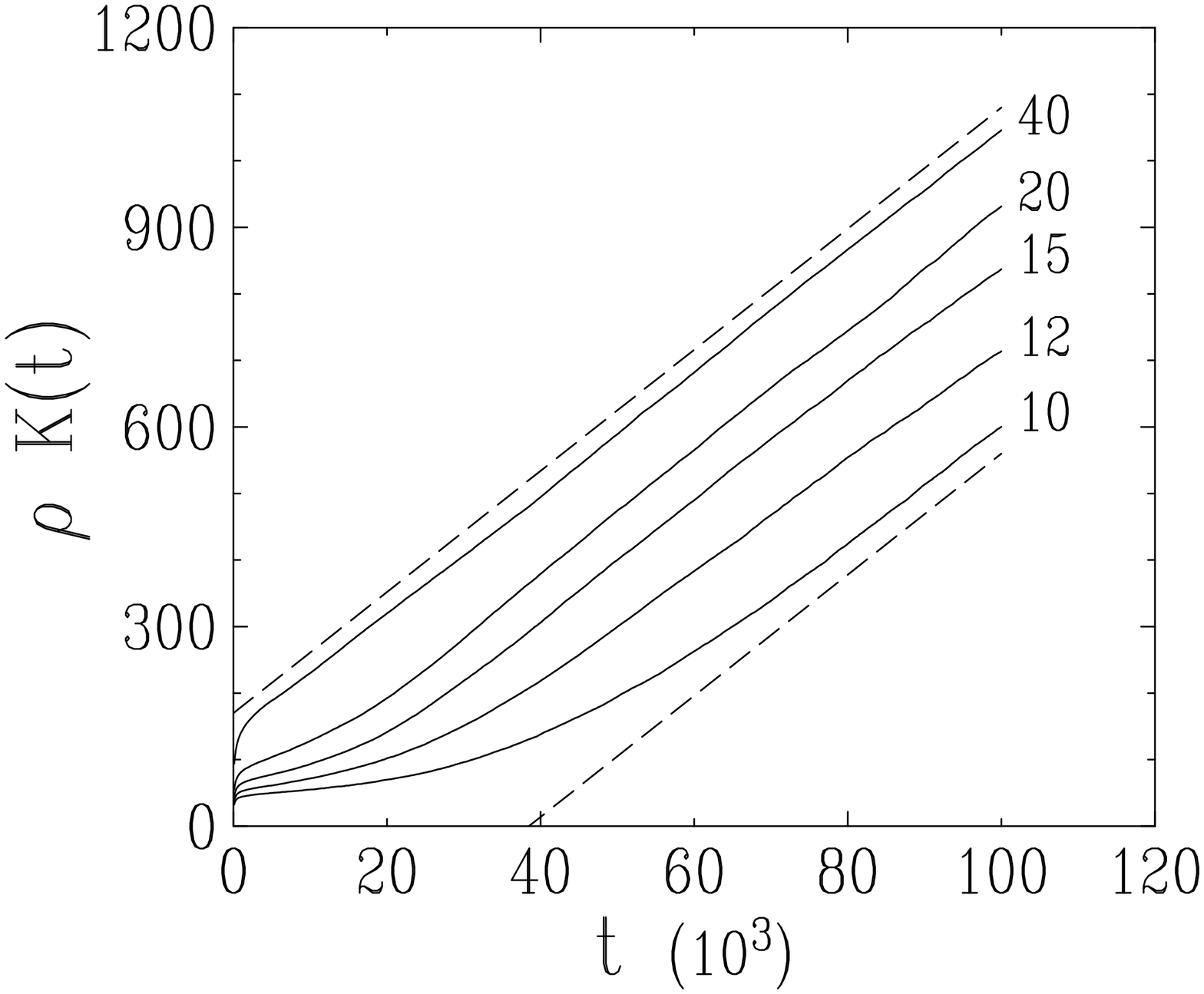}
\caption{\small
Plot of the reduced second moment $K(t)$ of the occupation distribution
in the two-dimensional target process against time $t$,
for various values of the density $\rho$, indicated on the curves.
Left: data for moderate values of time.
Right: data for longer values of time, multiplied by density $\rho$.
The parallel straight dashed lines, meant as a guide to the eye,
have a slope $9.1\times10^{-3}$.}
\label{fig2kt}
\end{center}
\end{figure}

In analogy with the one-dimensional case, we begin with the transient dynamics,
starting from a random initial condition.
We monitor the local relaxation by means of the reduced second moment $K(t)$
of the occupation distribution, introduced in~(\ref{kdef}).
Figure~\ref{fig2kt} shows a plot of $K(t)$ against time~$t$.
For all values of the density~$\rho$, $K(t)$ is an increasing function of time,
starting from its initial value $K(0)=(2\rho+1)/\rho$.
The most significant feature to be observed on the data is the following:
$K(t)$ saturates to a finite limiting value, denoted $K$,
for values of the density up to the threshold value $\rho_0\approx6$.
We have~$K_0\approx4.8$ for $\rho=\rho_0$.
To the contrary, $K(t)$ grows indefinitely for larger values of the density.
The right panel of Figure~\ref{fig2kt} demonstrates that
the asymptotic growth law of~$K(t)$ for all~$\rho>\rho_0$ is of the form
\beq
K(t)\approx\frac{Ct}{\rho},
\label{asylaw}
\eeq
where $C\approx9\times10^{-3}$ is taken from the slope
of the parallel dashed lines.

\begin{figure}[!tb]
\begin{center}
\includegraphics[angle=90,height=6truecm]{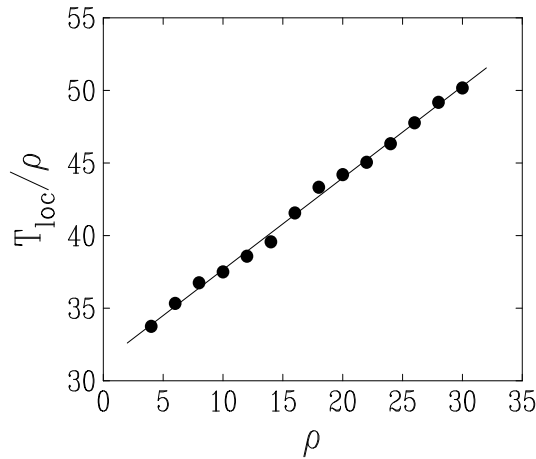}
\caption{\small
Plot of the local relaxation time $T_\loc$, divided by $\rho$,
against density~$\rho$.
Full straight line: least-squares fit with slope $A\approx0.63$.}
\label{fig2tloc}
\end{center}
\end{figure}

Pursuing along the lines of our investigation of the one-dimensional case,
we introduce a local relaxation time~$T_\loc$,
defined by the condition $K(T_\loc)=4$.
Figure~\ref{fig2tloc} shows a plot of $T_\loc$ so defined,
divided by~$\rho$, against~$\rho$.
The least-squares fit suggests a growth of the form~(\ref{tlinquad}),
with $A\approx0.63$, in a wide range of values of the density.
Notice that the local relaxation time exhibits no visible singularity
at the threshold density $\rho_0$.

\begin{figure}[!tb]
\begin{center}
\includegraphics[angle=90,height=6truecm]{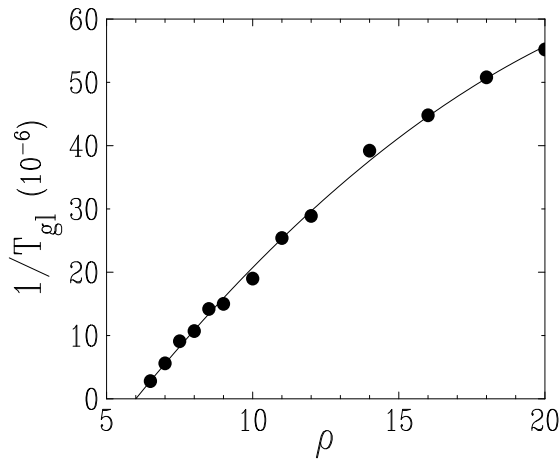}
\caption{\small
Plot of the reciprocal of the global time $T_\gl$ against density $\rho$.
Full line: second-degree polynomial fit yielding $\rho_0=6.0\pm0.1$.}
\label{fig2tg}
\end{center}
\end{figure}

We now turn to a more accurate determination of the threshold density $\rho_0$,
using the pre-asymptotic growth of $K(t)$ in the intermediate time regime,
where $K(t)$ has already departed in a significant way
from its plateau value~$K_0$,
but not yet reached the asymptotic linear growth~(\ref{asylaw}).
The right panel of Figure~\ref{fig2kt} demonstrates
that this regime lasts longer and longer
as the threshold density $\rho_0$ is approached.
This observation is turned to a quantitative measurement
by defining the global relaxation time $T_\gl$
by the condition $K(T_\gl)=K_0+\Delta K$,
where we set $K_0=4.8$, whereas the choice $\Delta K=50/\rho$
incorporates the form~(\ref{asylaw}) of the asymptotic growth law.
Figure~\ref{fig2tg} shows a plot of the reciprocal of the global time $T_\gl$
thus defined, against density.
The data convincingly demonstrate that the global time diverges at some
non-trivial threshold density~$\rho_0$.
A crossover of the data toward another type of asymptotic behavior
indeed seems extremely improbable,
in view of the accuracy of the available data.
The second-degree polynomial fit to the data shown on the plot
provides a rather accurate determination of the threshold density,
\beq
\rho_0=6.0\pm0.1,
\label{rhoc}
\eeq
as well as an evidence that the global time
diverges linearly as the threshold density is approached from above, as
\beq
T_\gl\approx T_0\;\frac{\rho_0}{\rho-\rho_0},
\label{tgldiv}
\eeq
with a rather large prefactor $T_0\approx28\,000$.

\begin{figure}[!tb]
\begin{center}
\includegraphics[angle=90,height=6truecm]{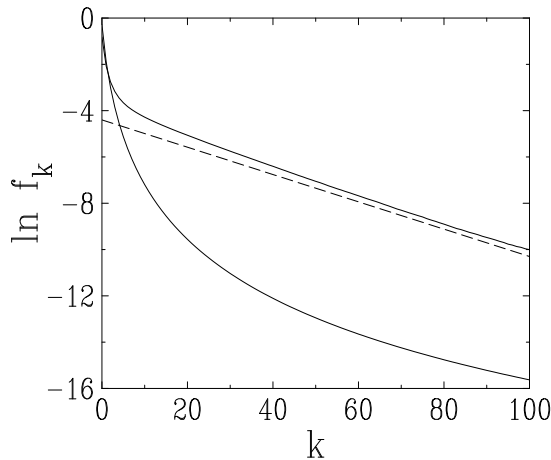}
\caption{\small
Logarithmic plot of occupation distributions.
Upper curve: fully asymmetric target process in two dimensions at its threshold
density $\rho_0$.
The dashed line, meant as a guide to the eye, has a slope $-\mu\approx-0.06$.
Lower curve: critical occupation distribution of the dual ZRP,
given by~(\ref{fkc}).}
\label{fig2f}
\end{center}
\end{figure}

The threshold density $\rho_0$ is the maximal density for which a homogeneous
fluid phase is stable.
At variance with the critical density $\rho_c$ of the symmetric target process
and of the dual ZRP, $\rho_0$ rather appears as a dynamical threshold.
This viewpoint is corroborated by the fact
that the stationary state at density~$\rho_0$ does not exhibit
any critical feature.
The distribution of the site occupations at the threshold density,
shown in Figure~\ref{fig2f}, has an exponential fall-off
of the form $f_k\sim\exp(-\mu k)$, with $\mu\approx0.06$,
at least in the accessible range of values of the occupation.
We checked that the data are not affected
in an appreciable way by finite-size effects in the range considered.
The critical occupation distribution~(\ref{fkc}) of the dual ZRP
is shown on the same plot as a comparison.
The latter distribution has a much smaller density $\rho_c=1/2$,
some 12 times smaller than the observed threshold density~(\ref{rhoc})
of the asymmetric two-dimensional model, but a slower power-law fall-off,
so that the distributions eventually cross each other.

\begin{figure}[!tb]
\begin{center}
\includegraphics[angle=90,height=6truecm]{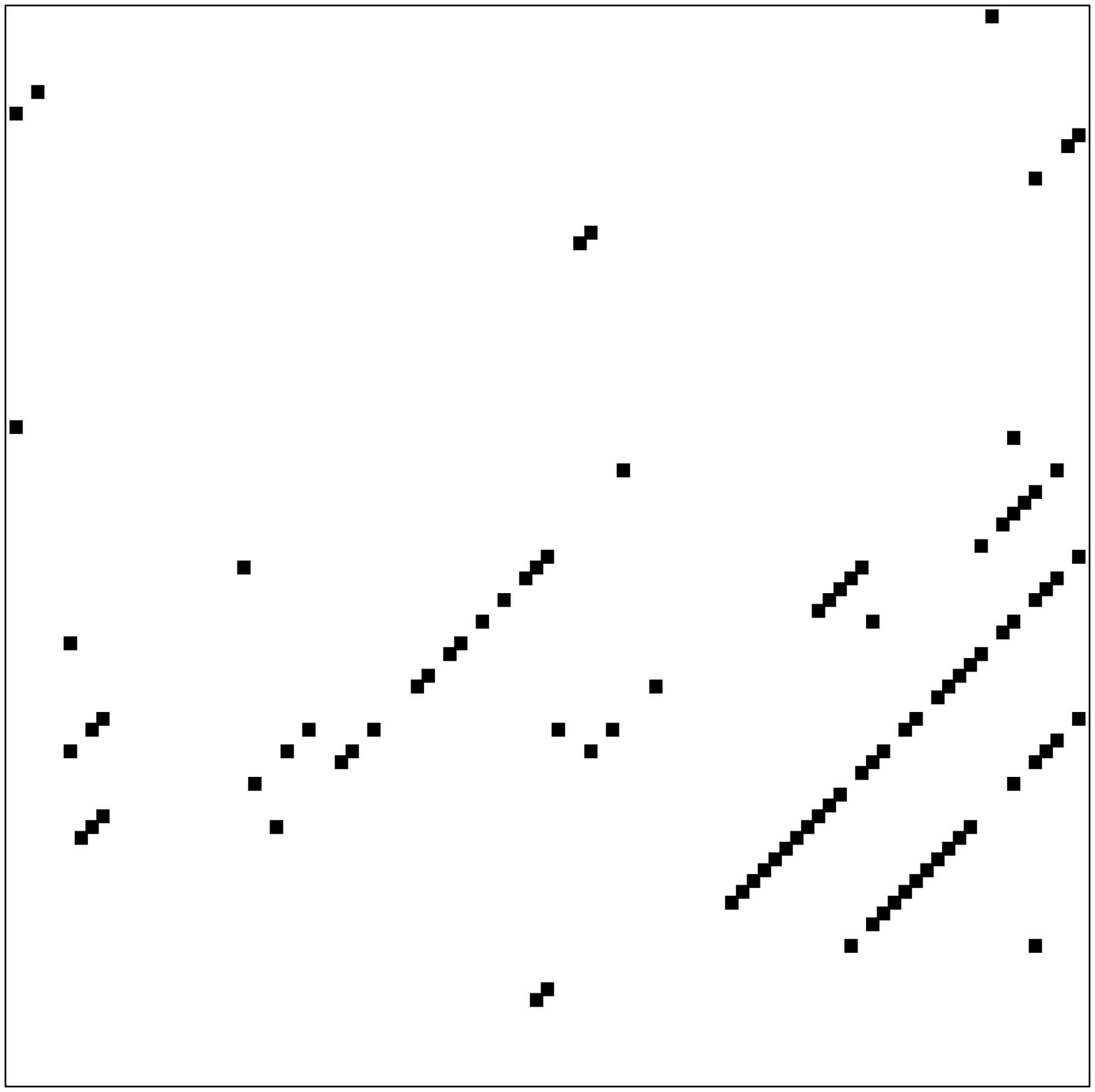}
{\hskip 15pt}
\includegraphics[angle=90,height=6truecm]{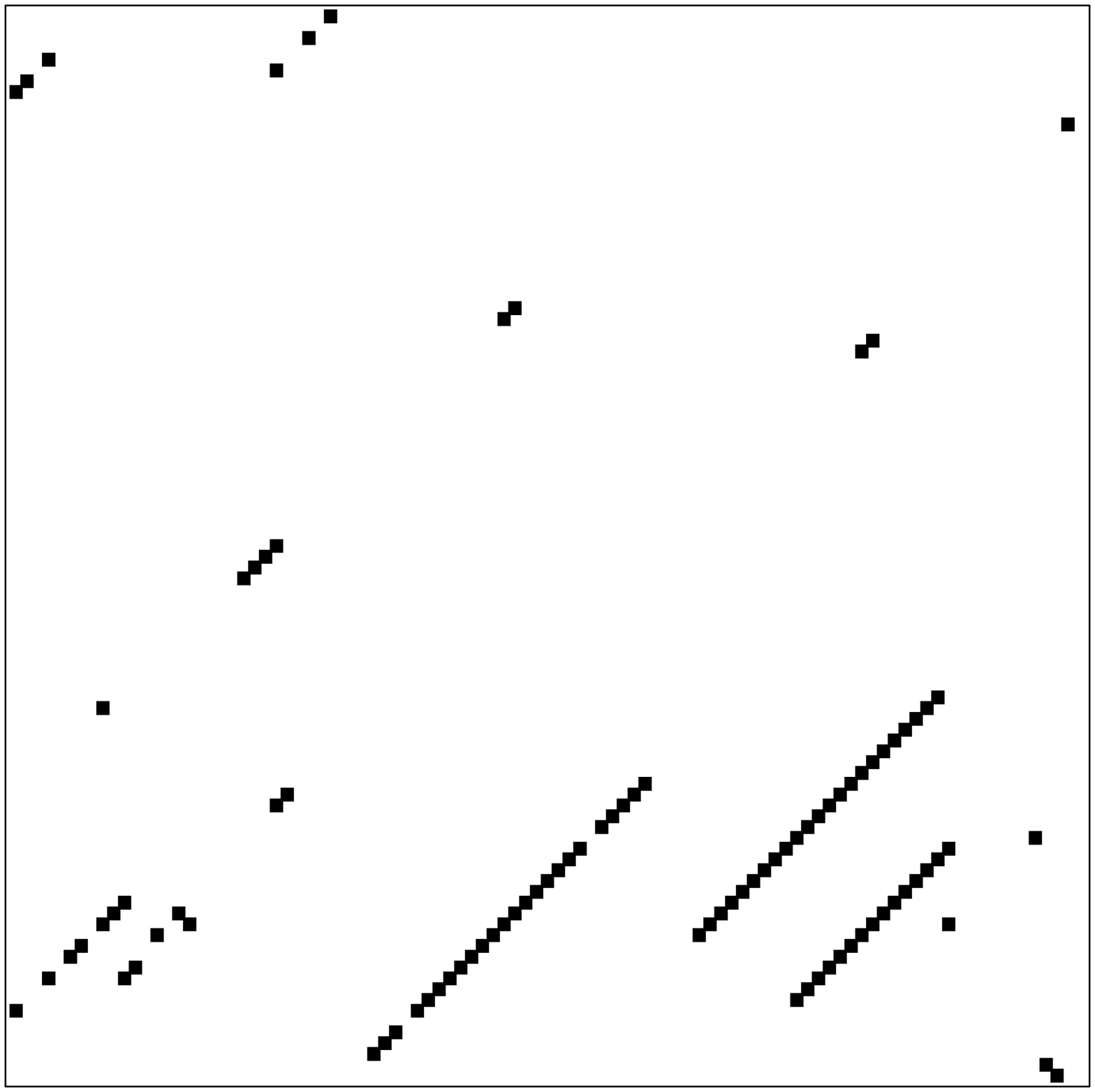}
\caption{\small
Plots of the 100 (i.e.,~1\%) most occupied sites
of a sample of size $100\times100$ at density $\rho=20$
in the coarsening regime.
Left: $t=5\times10^4$.
Right: $t=10^5$.}
\label{fig2plot}
\end{center}
\end{figure}

The progressive emergence of highly occupied coherent structures,
which are strongly elongated along the direction of the bias,
is illustrated in Figure~\ref{fig2plot},
showing two snapshots of the coarsening regime
of a sample of size $100\times100$ at density $\rho=20$.
The filled symbols show the 1\% most occupied sites.
The visible structures clearly are precursors of the extended condensate
mentioned above.
The background density of the fluid phase besides these structures
is found to be much smaller than the threshold density~$\rho_0$,
and comparable to the critical density $\rho_c=1/2$ of the dual ZRP.
We shall return to this point in more detail below (see Figure~\ref{figrhobg}).

\begin{figure}[!tb]
\begin{center}
\includegraphics[angle=90,height=6truecm]{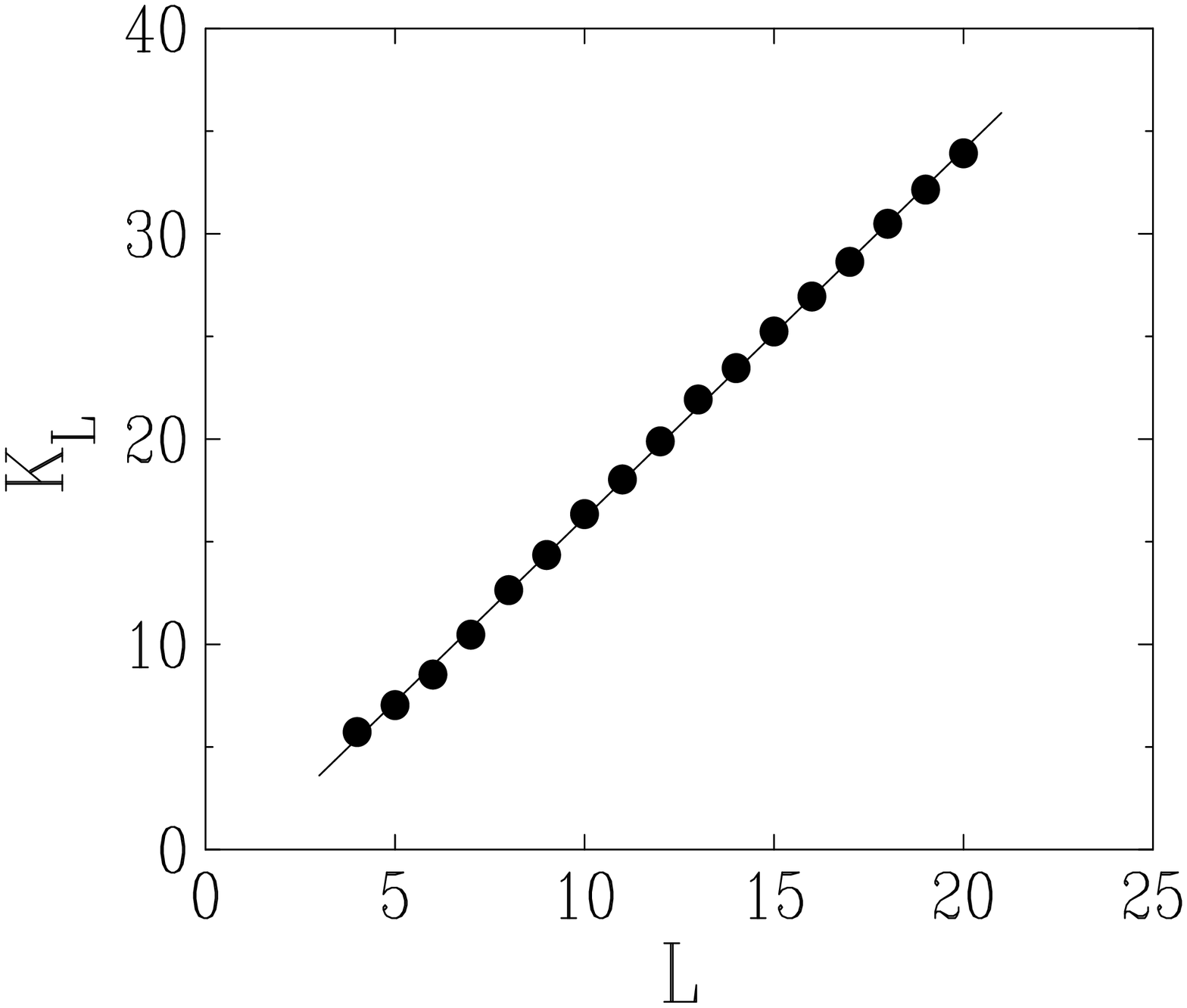}
{\hskip 15pt}
\includegraphics[angle=90,height=6truecm]{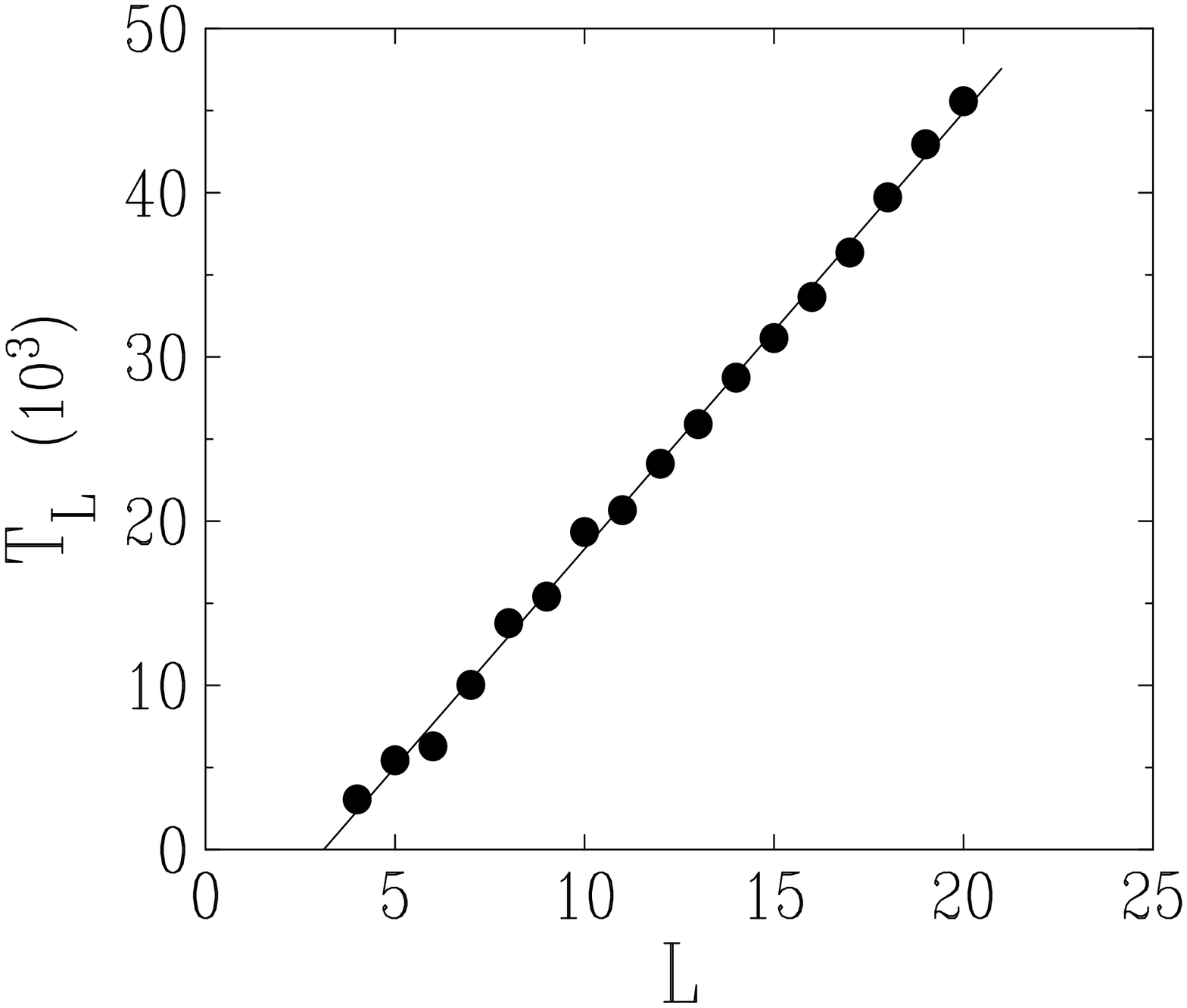}
\caption{\small
Plots of data concerning the stationary state of finite systems
at fixed density $\rho=20$, against their linear size $L$.
Left: stationary-state value $K_L$ of the reduced second moment
of the occupation distribution.
Right: relaxation time $T_L$.
Full straight lines: least-squares fits with respective slopes 1.79 and
2\,660.}
\label{figfss}
\end{center}
\end{figure}

\begin{figure}[!tb]
\begin{center}
\includegraphics[angle=90,height=6truecm]{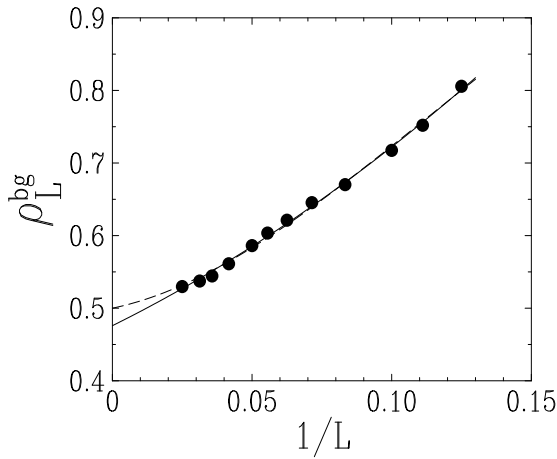}
\caption{\small
Plot of the background density $\rho_L^\bg$
in the stationary state of finite systems
at fixed density $\rho=20$, against their reciprocal linear size $1/L$.
Full line: second-degree polynomial fit yielding the extrapolated
value $\rho^\bg\approx0.48$.
Dashed line: constrained polynomial fit of degree 3 imposing
$\rho^\bg=\rho_c=1/2$.}
\label{figrhobg}
\end{center}
\end{figure}

We now study the typical characteristic sizes (width and height)
of the extended condensate in the stationary state of a finite system,
and of its precursors in the coarsening regime of an infinite system.
The width $W$ of a condensate is defined as the number of sites
which take part in the condensate,
whereas its height $H$ is the mean number of particles per site
in the condensate, so that the product~$HW$
gives a measure of the number of particles involved in the condensate.
We first consider the coarsening regime of an infinite system.
The condensate precursors shown in Figure~\ref{fig2plot}
are expected to be characterized by a typical
width $W(t)$ and height $H(t)$, with both scales growing with time.
The contribution of these precursors
to the reduced second moment of the occupation distribution
can be checked to scale as $K(t)\sim H(t)/\rho$,
irrespective of the width $W(t)$.
The growth law~(\ref{asylaw}) therefore implies
that the height of condensate precursors grows linearly in time, according to
\beq
H(t)\approx Ct,
\label{ht}
\eeq
with $C\approx9\times10^{-3}$.
This asymptotic coarsening law is expected to hold
for any density~$\rho>\rho_0$.
The behavior of the width $W$ of condensate precursors can only be investigated
in an indirect way, by means of finite-size scaling.
We therefore consider finite systems,
namely square samples of linear size $L$, with periodic boundary conditions.
Figure~\ref{figfss} shows numerical data concerning
the stationary state of finite systems against their linear size~$L$,
at fixed density $\rho=20$, well above the threshold density~$\rho_0$.
The left panel shows the stationary-state value $K_L$
of the reduced second moment of the occupation distribution.
The right panel shows the characteristic relaxation time $T_L$,
defined by the condition $K(T_L)=(K_L+K_0)/2$, again with $K_0=4.8$.
The data for both quantities clearly exhibit a linear growth with the size $L$.
The first of these growth laws implies $H_L\approx\rho K_L\sim L$.
This is in accord with the expectation
that typical stationary-state configurations
have a single and roughly system-spanning extended condensate,
for which $H_L\sim L$.
Furthermore, as the number of particles involved in the condensate
scales as $H_LW_L\sim L^2$, we have $H_L\sim W_L\sim L$.
Let us now make the finite-size scaling assumption
that $H(t)$ and $W(t)$ become respectively comparable to $H_L$ and $W_L$
for a time $t$ comparable to the relaxation time $T_L$.
This yields the scaling law $H(t)\sim t$,
already known~(see~(\ref{ht})), and the prediction $W(t)\sim t$.

Finally, we have also measured the background density $\rho_L^\bg$ of the fluid
phase in the stationary state of finite samples at density $\rho=20$.
This quantity is algorithmically defined as follows.
For any intercept $k=1,\dots,L$, consider
the total number of particles in the diagonal array
with intercept $k$, i.e., with equation $n=m+k\ (\hbox{mod.}\ L)$:
\beq
\N_k=\sum_{m=1}^LN_{m,m+k}.
\eeq
The largest of these $L$ numbers, $\N_\max$,
corresponds to the diagonal array occupied by the extended condensate.
It is overwhelmingly larger than the others, as it scales as $\N_\max\sim L^2$.
The other $(L-1)$ numbers $\N_k$ represent the fluid phase,
and therefore scale as $\rho_L^\bg L$.
We are thus naturally led to define the background density as
\beq
\rho_L^\bg=\frac{N-\N_\max}{L(L-1)},
\eeq
where $N$ is the total number of particles
in the system.
Figure~\ref{figrhobg} shows a plot of the stationary-state
background density $\rho_L^\bg$ for finite samples of size $L$, against $1/L$.
The data demonstrate that $\rho_L^\bg$ is smaller than unity,
and therefore much smaller than the mean density $\rho=20$, as soon as $L\ge7$.
The segregation phenomenon is therefore already fully at work
for rather small system sizes.
From a quantitative viewpoint, a second-degree polynomial fit to the data yields
the extrapolated value $\rho^\bg\approx0.48$.
The limiting value thus obtained is remarkable close
to the critical density $\rho_c=1/2$ of the dual ZRP.
Imposing the constraint $\rho^\bg=\rho_c=1/2$ indeed hardly changes
the fit (dashed line).
This agreement strongly suggests that the background fluid phase
of the two-dimensional target process above its threshold density
is characterized by the critical occupation distribution of the dual ZRP,
just as the S-sites of the one-dimensional case in the high-density limit.
As a consequence, the density $\rho^\bg$ of the fluid phase:
\beq
\rho^\bg=\left\{\matrix{
\rho\hfill&\hbox{for}\hfill&\rho<\rho_0,\cr
\rho_c\hfill&\hbox{for}\hfill&\rho>\rho_0,}\right.
\label{rhojump}
\eeq
has a discontinuous jump at the threshold density $\rho_0$.

\section{Discussion}
\label{quatre}

In this work we introduced a novel example
of a migration process, the target process.
We then studied in detail the structure of the nonequilibrium stationary state
of the asymmetric target process, the main focus being on
the fate of the condensation phenomenon.

The stationary-state measures of migration processes
do not have a product form in general.
The {\it symmetric} target process, though,
has the same stationary-state product measure as the corresponding dual ZRP.
In particular, the so-called canonical target process, defined by the
rate~(\ref{vcan}) dual to the ZRP with rate~(\ref{ucan}),
has a continuous condensation transition
at a finite critical density $\rho_c$ whenever $b>2$,
with a macroscopic condensate occupying a single site
for densities $\rho>\rho_c$, irrespective of the dimensionality of the system.
The {\it asymmetric} target process has a fluctuating stationary state
with non-trivial spatial and temporal correlations,
whose qualitative features depend on the dimensionality.
Our main effort in the present work consisted in characterizing
this nonequilibrium stationary state,
including its dependence on the dimensionality.

We have reached a complete understanding of the one-dimensional target process,
especially in the high-density regime of most interest.
We showed, by exploiting the existence of a conserved current,
that the asymmetric canonical target process has no condensation transition,
and remains homogeneous at any finite density.
In the high-density regime, an alternating scenario prevails for $b>1$:
typical configurations consist of long alternating sequences BSBSBSBS...
of highly occupied B-sites, and less occupied S-sites, whose occupation
distribution coincides with the critical distribution of the dual ZRP.
The coherence length (mean domain size) of this alternating
structure diverges as $\xi\sim\rho^2$.
We also gave a characterization of the scaling behavior of many other
quantities in the vicinity of the `infinite-density fixed point'.

For the asymmetric target process in higher dimensions,
we argued that the condensate must be extended
and have the form of a one-dimensional structure
elongated along the direction of the bias.
In the two-dimensional case, numerical simulations performed for $b=4$
show that the model exhibits an unconventional condensation transition
at the density $\rho_0\approx6$.
This density, which is much larger than the critical density $\rho_c=1/2$
of the dual ZRP, appears as a dynamical threshold:
it is the maximal density at which a homogeneous fluid phase
is dynamically stable.
For $\rho>\rho_0$, the predicted extended condensate is observed,
whereas the background fluid phase again appears as critical.
This picture seems to be generic for higher-dimensional systems.
Preliminary numerical simulations
of the asymmetric target process on the three-dimensional cubic lattice
(where the displacement is along either of the unit vectors
$\bfe_1$, $\bfe_2$, $\bfe_3$ with equal probabilities),
indeed show that the overall picture is quite similar
to the two-dimensional one.
The global relaxation time~$T_\gl$ is again found
to diverge according to~(\ref{tgldiv}), with $\rho_0\approx34$ for $b=4$.

Let us mention that another mechanism
leading to an extended condensate in a class of mass transport models
in one dimension has been reported recently~\cite{ehm}.
There, the nonequilibrium stationary-state measure is a product
whose factors involve the occupations of two consecutive sites.
In the condensed phase, those models exhibit an extended condensate,
whose height and width scale as $H_N\sim W_N\sim N^{1/2}$
for a finite system of $N$ sites.
These scaling laws are formally identical to those found in the present work.

The existence of a threshold density $\rho_0$ at which
the background density has a discontinuous jump (see~(\ref{rhojump}))
is reminiscent of what occurs in the model studied in~\cite{zrp2},
namely a ZRP with two species of particles, and with rates such that
the stationary-state measure does not have a product form.
When the densities $\rho^{(1)}$ and~$\rho^{(2)}$ of the two species are equal,
the behavior of the system is qualitatively the same as that
of the canonical ZRP (with one species).
In particular the system has a continuous phase transition
at some critical density $\rho_c$, from a fluid phase to a
condensed phase with critical background.
The general situation where the two densities are different
however drastically departs from this known scenario.
If either of the two densities ($\rho^{(1)}$, say) is kept fixed
at a value larger than $\rho_c$, on increasing the other density $\rho^{(2)}$,
the system remains homogeneous as long as $\rho^{(2)}$ is less than a
threshold value $\rho^{(2)}_0$ which depends on~$\rho^{(1)}$.
At this threshold the system undergoes a discontinuous transition
from an imbalanced fluid phase,
where both species have densities $\rho^{(1)}$ and $\rho^{(2)}_0$
larger than the critical density, to an imbalanced condensate
coexisting with a balanced critical fluid
with densities $\rho^{(1)}=\rho^{(2)}=\rho_c$.

Finally, the observed rapid growth of the threshold density
$\rho_0$ with the di\-men\-sio\-na\-li\-ty raises the question of the behavior
of the asymmetric target process in high dimensions.
At this point let us emphasize that the absence of a stationary-state product
measure for the target process
is a rather subtle effect which needs the conjunction of several ingredients,
and chiefly the presence of a bias.
This feature cannot be present in mean-field geometries
such as the complete graph,
so that the dynamical threshold behavior of the model in high dimensions
is not expected to smoothly converge to a well-defined mean-field limit.

\end{document}